\documentclass[pre,twocolumn,superscriptaddress,showpacs,floatfix]{revtex4}

\usepackage{graphicx}

\begin{document}

\title{Statistical physics of cerebral embolization leading to stroke}

\author{J. P. Hague}
\affiliation{Department of Physics and Astronomy, The Open University, MK7 6AA, UK}

\author{E. M. L. Chung} \affiliation{Department of Cardiovascular Sciences,
University of Leicester, LE1 5WW, UK}

\date{\today}

\begin{abstract}

We discuss the physics of embolic stroke using a minimal model of
emboli moving through the cerebral arteries.  Our model of the blood
flow network consists of a bifurcating tree, into which we introduce
particles (emboli) that halt flow on reaching a node of similar
size. Flow is weighted away from blocked arteries, inducing an
effective interaction between emboli. We justify the form of the flow
weighting using a steady flow (Poiseuille) analysis and a more
complicated nonlinear analysis. We discuss free flowing and heavily
congested limits and examine the transition from free flow to
congestion using numerics. The correlation time is found to increase
significantly at a critical value, and a finite size scaling is
carried out. An order parameter
for non-equilibrium critical behavior is identified as the overlap of
blockages' flow shadows. Our work shows embolic stroke to be a feature
of the cerebral blood flow network on the verge of a phase transition.

%
%


\end{abstract}

\maketitle

\section{Introduction}

Stroke is a common cause of death and disability, and is expected to become more prevalent as average life expectancy increases \cite{lopez1998a}. A common cause of stroke is from embolisation, where small pieces of thrombus and plaque detach from the insides of diseased arteries and move through the vasculature to become lodged in arteries supplying the brain. Depending on the duration and position of the obstruction, embolic blockages can prove fatal.

Doppler ultrasound embolus detection in individuals at high risk of stroke reveals that large numbers of emboli typically enter the cerebral circulation in advance of a major stroke occurring \cite{mackinnon2005a}. Embolus detection
and autopsy studies of patients who have undergone cardiovascular surgery
 show that
patients can experience several thousand emboli during typical heart
surgery \cite{moody1990a} (which carries a 3-10\% stroke risk \cite{bucerius2003a}). The interplay between these emboli is
governed by flow dynamics, and thus the onset of stroke can be considered as a complex process of arterial obstruction involving multiple blockages. However, little is currently understood about the impact of multiple emboli leading to the onset of stroke.

To gain a better understanding of the triggers leading to the onset of
stroke from multiple emboli, we developed a probabilistic Monte Carlo approach to
predict the severity of arterial obstruction \cite{chung2007a}. Our model is non-trivial because blockages
divert the flow of new emboli that enter the arteries (i.e. there is an effective
interaction between emboli). We predicted a very rapid change from
free-flow with little occlusion to severely obstructed arteries as the
size of the emboli increased, which we suspected was associated with a
phase transition. In this article, we examine the origins of this
behavior more closely.

This article is organized as follows. In section \ref{sec:model}, we
introduce our model and algorithm, and carry out a basic fluid
dynamical analysis of the vascular tree to justify our flow weighting
scheme. To gain more insight into the model, we determine the limiting
properties of the model in section \ref{sec:limits}.
To
better understand the motion of emboli through the tree, we examine
space-time plots of blockages in a single realization of the model (section \ref{sec:spacetime}). We examine the time correlator and
correlation time in Sec. \ref{sec:timecorrelator}. The order parameter is
examined in Sec. \ref{sec:orderparameter}. Finally, we summarize our results in
section \ref{sec:summary}.

\section{Model}
\label{sec:model}

\subsection{Vasculature}

\begin{figure}
\includegraphics[width = 78mm]{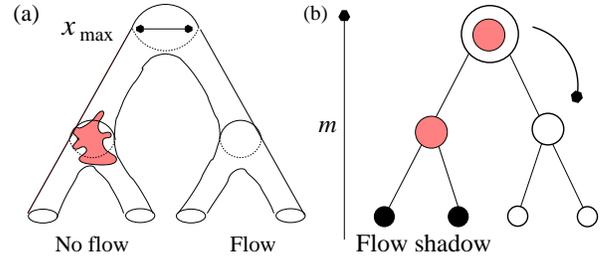}
\caption{A representation of a small section of the cerebral vasculature. (a) Arteries
supplying the brain typically bifurcate. The number of branches in the
vasculature is prohibitively large for a full fluid-dynamical
analysis. For this reason, we develop a minimal model (b) where nodes
on a bifurcating tree represent arteries which can become
blocked. When emboli block nodes, blood flow is weighted according to
the number of nodes at the bottom of the tree receiving unimpeded flow. The
flow carries new emboli away from existing blockages, introducing
an effective interaction between emboli.}
\label{fig:model}
\end{figure}

We treat the cerebral vasculature as a bifurcating tree, which is
consistent with high resolution images showing that $>97\%$ of all
branches in the cerebral arteries are bifurcations
\cite{cassot2006a}. In this study, we do not consider the largest
vessels such as the circle of Willis (CoW). Since most emboli travel
into the middle carotid artery (MCA) after passing through the CoW, we
consider our model to be a simplified version of the cerebral arteries
supplied by the MCA. Bifurcations can be characterized by the
bifurcation exponent, which is defined through the relation
$x_s^{\gamma} = 2 x_d^{\gamma}$ where $x_s$ and $x_d$ are the
respective radii of the source and daughter vessels
\cite{karshafian2003a}. In this study, we use a single bifurcation
exponent of $\gamma$ at all levels of the tree. When the bifurcation
exponent is greater than $2$, the total cross sectional area of the
daughter vessels is greater than that of the source vessel. In many
organisms, the bifurcation exponent is expected to be approximately $3$
\cite{murray1926a}. A schematic of our model vasculature is shown in
Fig. \ref{fig:model}. We vary $M_{\ell}$ (the total number of layers
in our tree) so that the vessel sizes range from $x_{\rm max}=1$mm to
a few $\mu$m ($x_{\rm min}$), which is similar to the dimensions of the
arteries in the brain. Node diameters are $x_{m}=x_{\rm
min}2^{m/\gamma}$ with $m$ the level of the tree ($m=0$ corresponds to
the smallest arteries and $m = M_{\ell} - 1$ to the largest). The
nodes are connected by segments representing arteries.

\subsection{Fluid dynamics and flow weighting}

\begin{figure}
\includegraphics[height=85mm,angle=270]{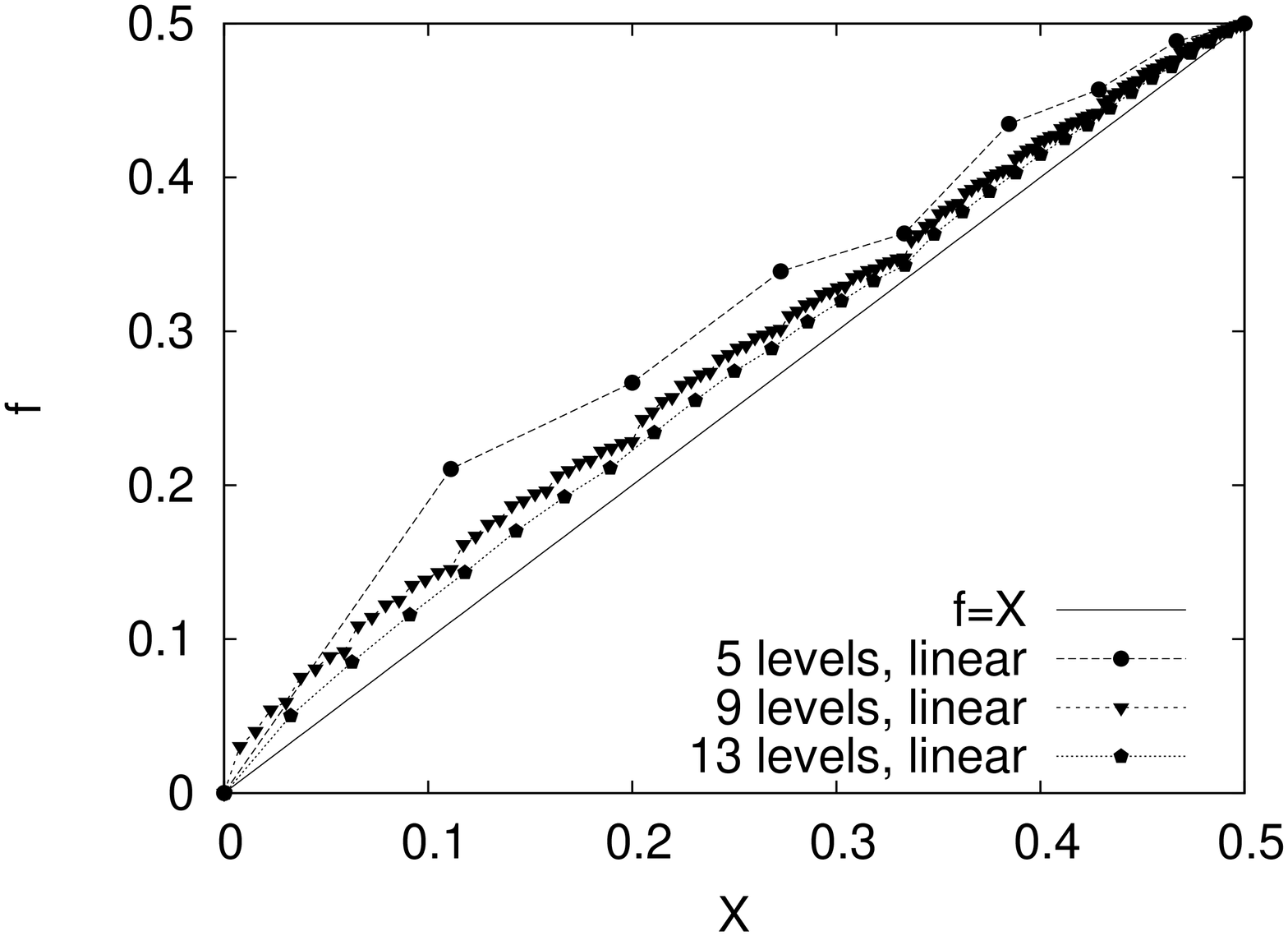}
\includegraphics[height=85mm,angle=270]{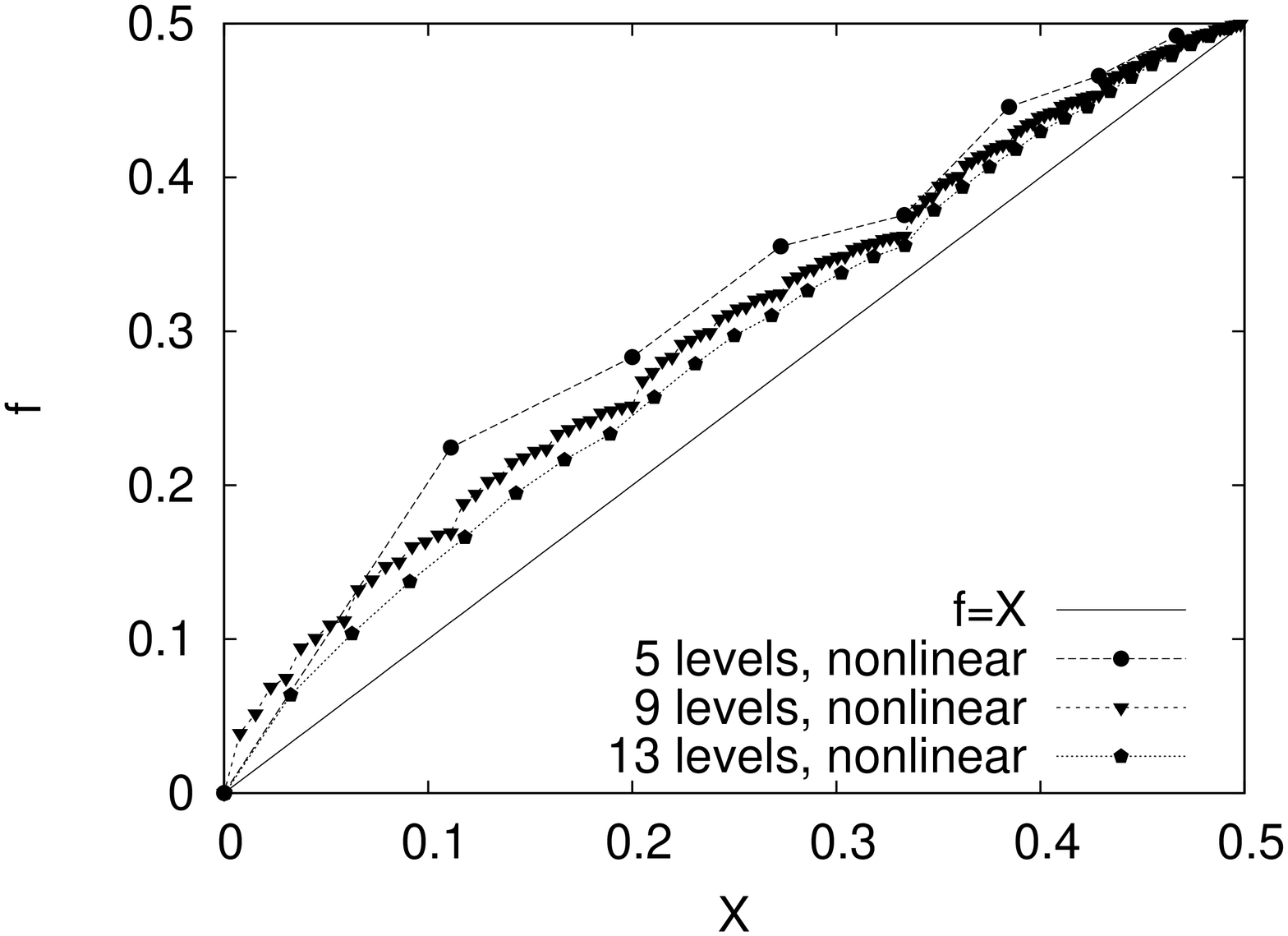}
\caption{(a) Flow weighting on small symmetric trees where the drop in pressure
obeys Poiseuille's equation, and the bifurcation exponent $\gamma=3$. Flow resistances were assumed
to change with tree level according to Eqn. \ref{eqn:resistance}. (b) Flow weighting on small symmetric trees where pressure
differences have a nonlinear dependence on the flow according to Eqn. \ref{eqn:nonlinear} and $\gamma=3$. The flow weightings tend towards the linear weighting scheme with increasing system size.}
\label{fig:resistormodel}
\end{figure}

Emboli are carried by the blood, and as will be discussed in sec. \ref{sec:trajectory}, the trajectory of emboli at a bifurcation is related to the relative flow in the
branches. Since obstruction by emboli causes
redirection of blood-flow, it is relevant to determine the ratio of flows in the network downstream from a bifurcation in the presence of blockages using a basic fluid dynamical analysis.

For initial determination of how the flow is modified by blockages downstream of a bifurcation, it is convenient to assume
that flow in segments of the tree obeys Poiseuille's law,
\begin{equation}
\Delta P = \left( \frac{8\mu l}{\pi a^4} \right) q \equiv R q
\end{equation}
where $\mu$ is the blood viscosity, $a$ is the radius of the artery,
$l$ is the length of the segment of the artery that is being
considered, $q$ is the volume flow through the vessel, $R$ is the
resistance of the vessel to flow, and $\Delta P$ is the pressure drop
over a vessel segment. By following all possible routes from the root
node where the pressure is $P_{A}$ to the capillary mesh where
pressure is $P_{V}$, a set of $2^{N-1}$ equations can be determined,
\begin{equation}
\sum_{n=0}^{N-1} R_{n,i(n)} q_{n,i(n)} = P_{A}-P_{V}
\end{equation}
which are supplemented by $2^{N-1}-1$ simultaneous equations of the form
\begin{equation}
q_{n,i} - q_{n-1,2i-1} - q_{n-1,2i} = 0
\end{equation}
derived from conservation of flow to give $2^{N}-1$ equations for the
$2^{N}-1$ flows in the tree. The nomenclature $q_{n,i}$ indicates the
$i$th node in the $n$th level.

It remains to assign realistic
values to the resistances. The lengths of arteries are proportional to
their radii $l=\kappa a$ \cite{west1997a} and the radii typically
follow Murray's law $a_{n-1}^{\gamma} = 2a_{n}^{\gamma}$ with the
bifurcation exponent normally $\gamma =3$ \cite{murray1926a} (note that in this section only, $n=0$ is the root node, with $n$ increasing on entering the tree). Thus
$R_n\propto a^{-3} \propto 2^{3n/\gamma}$. Replacing the coefficients,
\begin{equation}
R_n = \frac{8\mu\kappa 2^{3n/\gamma}}{\pi a_0^3}
\label{eqn:resistance}
\end{equation}
$a_0$ is the radius of the root node. Following standard theoretical practice, the ratio of segment length to segment radius is set as $l=20a$, i.e. $\kappa=20$ \cite{zamir1999a}.

It is straightforward to invert the simultaneous equations to
establish the proportion of total flow into a daughter vessel,
$f=q_{A}/(q_{A}+q_{B})$, the proportion of flow in direction $A$ at a
bifurcation as $X$ (the ratio of free nodes in direction $A$ to total free
nodes) is changed (the flow in daughter vessels $A$ and $B$ is
$q_{A}$ and $q_{B}$ respectively) \footnote{We carried out the
solution using GNU Octave.}. The result is shown in figure
\ref{fig:resistormodel}(a). As the tree size grows, the relative
resistance of the end vessels increases as compared to the root node,
and the flow solution approaches the line $f=X$. For $\gamma<3$ the
linear approximation is even more closely approached.

Directly downstream from a bifurcation, Poiseuille flow is not fully formed, leading to an increased pressure drop which depends nonlinearly on the flow according to the formula \cite{cieslicki2005a},
\begin{equation}
\Delta P = R\left(q+\frac{0.176q^2\rho}{l\pi\mu}\right),
\label{eqn:nonlinear}
\end{equation}
here $\rho$ the density of the fluid. We solve the set of equations
resulting from replacing Poiseuille's law for the nonlinear pressure
drop of equation \ref{eqn:nonlinear} (using the results from the
linear flow analysis as an initial solution) showing the results in
Fig. \ref{fig:resistormodel}(b). Again, the line $f=X$ is approached
as the size of the tree in increased. The approach is less rapid than
that found with the Poiseuille analysis, but for very large trees, a
linear relation between the ratio of free arterioles to the ratio of
flows is expected.

The results of the basic fluid dynamical analysis indicate that a
linear flow weighting scheme is a good approximation to the true fluid
dynamics. The linear weighting approximation is strictly valid in the
limit that the flow resistance in the capillary nodes is much higher
than in any of the other nodes. It is also known from studies of
vascular physiology that most pressure is dropped across the smallest
arterioles (see e.g. Ref. \cite{levick2003a}) giving further evidence
for the validity of a linear weighting approximation.

\subsection{Embolus trajectory at bifurcation}
\label{sec:trajectory}

\begin{figure}
\includegraphics[height=75mm,angle=270]{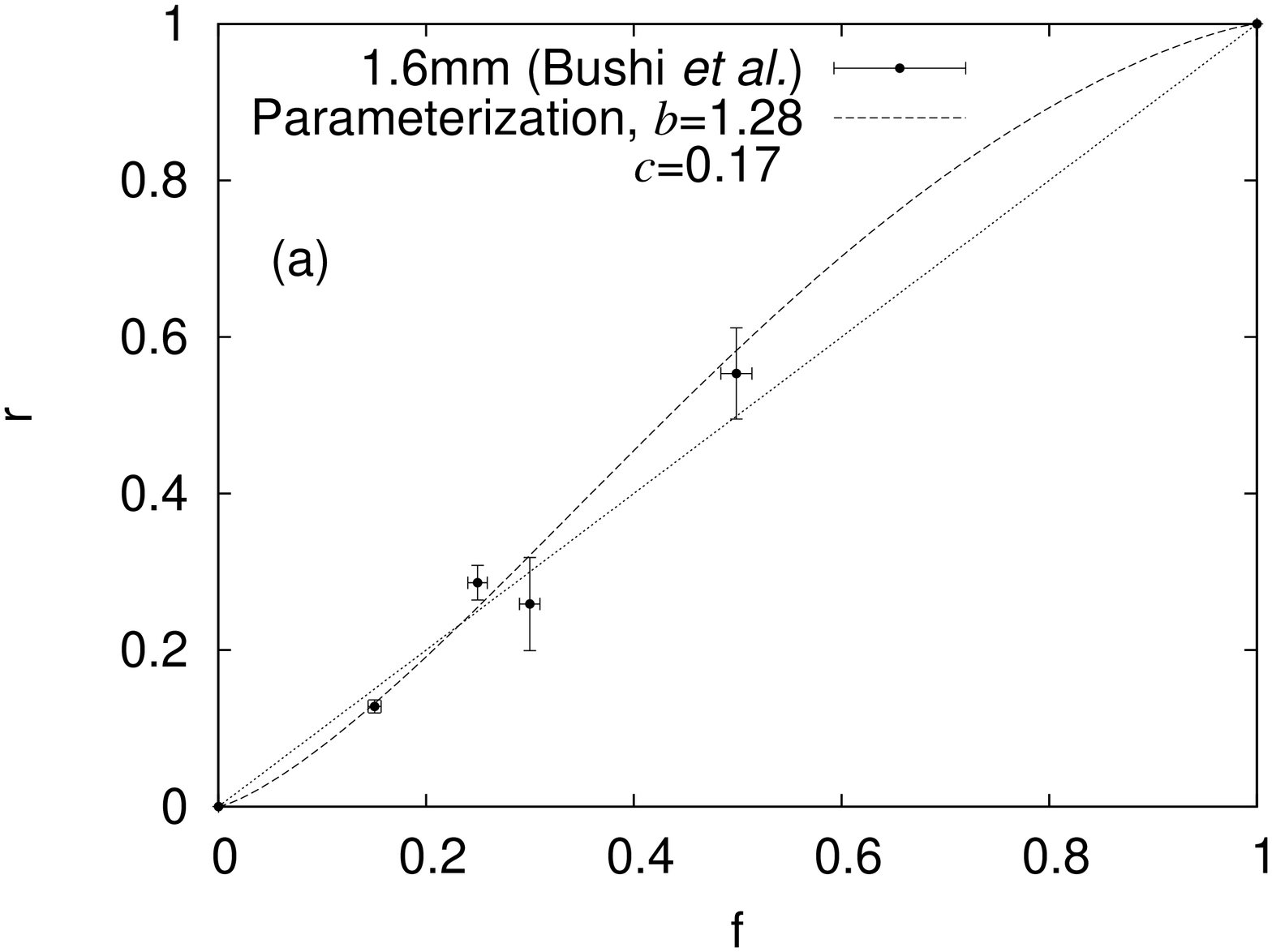}
\includegraphics[height=75mm,angle=270]{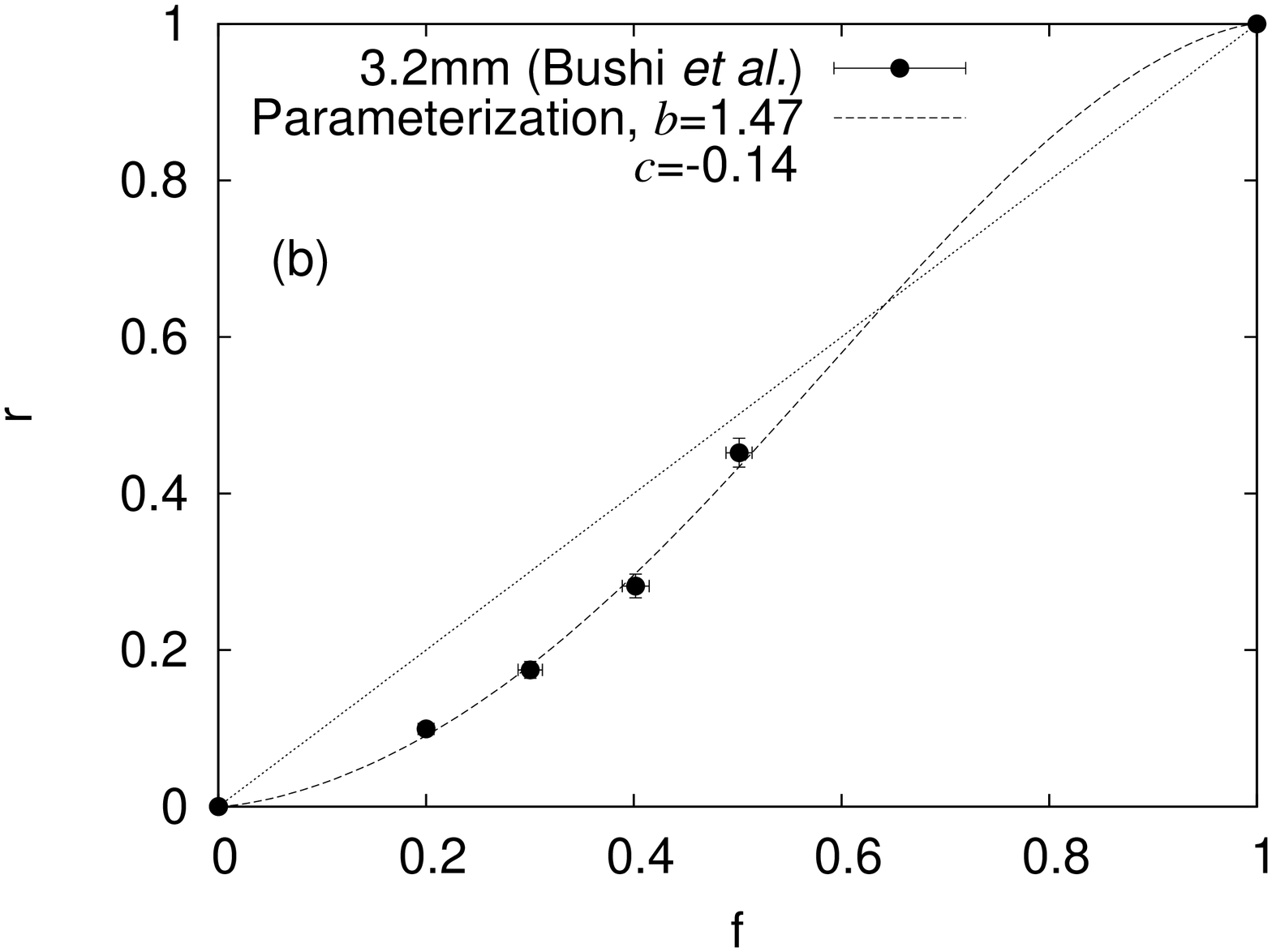}
\caption{$r$ vs $f$ at a bifurcation. Data shown as solid points are
taken from {\it in vitro} measurements by Bushi {\it et al}. \cite{bushi2005a}.
Our parameterization of the experimental data via equation \ref{eqn:leftrightprop} is
shown as the dotted line. }
\label{fig:bushi}
\end{figure}

The response of an embolus to flow at a bifurcation is complicated. In the
limit that the size of an embolus approaches that of the blood cells,
the ratio of the average number of emboli that travel into each of the
branches must be identical to the ratio of flows. However, when emboli
are large compared with the sizes of vessels, the proportion of
particles flowing into the daughter branches does not have a simple
form \cite{bushi2005a}.

Bushi {\it et al.} have performed experiments to establish the
trajectories of emboli at a single bifurcation, and in
Fig. \ref{fig:bushi} we reproduce some of their data from
Ref. \onlinecite{bushi2005a}. Fig. \ref{fig:bushi} shows
$r=E_{A}/(E_{A}+E_{B})$, which is the probability that an embolus
travels into branch $A$ vs $f=q_{A}/(q_{A}+q_{B})$, the proportion of
flow in direction $A$ at the bifurcation (the flow in daughter vessels
$A$ and $B$ is $q_{A}$ and $q_{B}$ respectively, and the number of
emboli traveling into branches $A$ and $B$ is $E_{A}$ and
$E_{B}$). Data shown as solid points are taken from {\it in vitro}
measurements by Bushi {\it et al.} using pulsatile flow
\cite{bushi2005a}, and the dotted lines are our parameterizations of
the data. We note that Bushi {\it et al.} measured $E_{B}/E_{A}$ and
$q_{B}/q_{A}$, which we have converted to $p$ and $f$ using
$E_{B}/E_{A}=1/r-1$ and $q_{B}/q_{A}=1/f-1$. The diameter of the pipes
forming the bifurcation in the experiment by Bushi {\it et al.} were
quite large (6mm for parent and daughter B, and 4mm for daughter A)
which explains why their `emboli' also needed to be large to see the
any nonlinear trajectory weighting \cite{bushi2005a}.

We suggest that the relation between probability that an embolus travels down branch $A$ and the proportion of flow in branch $A$ has the sigmoid-like form,
\begin{equation}
r = \frac{\tanh\left[b \tanh^{-1}(2f-1)+c\right]+1}{2}
\label{eqn:leftrightprop}
\end{equation}
where the parameters $b$ and $c$ depend on the embolus size
\footnote{We note that the choice of parameterization is not
unique. Any monotonic function which has two different constant asymptotes at
$x=\pm\infty$ and its inverse can be used in place of the $\tanh$,
e.g. the error function.}. This satisfies known limits, for example if
there is no flow ($f=0$), then no emboli can travel down the branch
($r=0$) and likewise if all the flow is in the branch ($f=1$), then
there is a probability $r=1$ that emboli flow into the
branch. Furthermore, the behavior of small emboli is known. When
$b=1$, the linear weighting is recovered, corresponding to very small
emboli (i.e. clots of a few blood cells) which generally follow the
flow of blood.  If branches are symmetric, then $f$ and $r$ are
symmetric through the map $f:=f'=1-f$ and $r:=r'=1-r$, which is why we
are confident of the sigmoid-like form. Thus for symmetric
bifurcations, the curve must go through the point $r=0.5, f=0.5$ and
$c=0$. Fits of equation \ref{eqn:leftrightprop} are also shown in
Fig. \ref{fig:bushi}. The parameters of our fits are $b=1.47\pm 0.07$,
$c=-0.137 \pm 0.025$ for the 3.2mm emboli and $b=1.28\pm 0.13$, $c=
0.170 \pm 0.097 $ for 1.6mm emboli. The small values of $c$ indicate
that embolus trajectory is more likely controlled by flow ratio than
the asymmetry of the vessels. The 1.6mm data could be compatible with
the line $r=f$, but the 3.2mm data are not. This indicates that for
emboli of diameters of the order of half the vessel diameter, the
linear weighting scheme is applicable.

We examine the effects of the more complex weighting scheme on the
crossover or phase transition by considering the extreme limits of the
parameterization with $b=1, c=0$ (which corresponds to a linear
trajectory weighting when emboli are small) and $b\rightarrow\infty,
c=0$, the latter corresponding a strong interaction limit where emboli
always travel in the direction of greatest flow. Note that currently
only symmetric bifurcations are considered.

Since embolization affects flow, the paths of subsequent emboli entering the
arterial tree are diverted away from existing blockages. This
generates an effective interaction between emboli, implying that the
onset of stroke should be considered as an interacting many-particle
problem.

\subsection{Algorithm}

We introduce `emboli' into the tree with size $d_0$, produced with
rate $\tau^{-1}$ (s$^{-1}$) and dissolving at a constant rate $\alpha
= \Delta d/\Delta t$ (mm/s) (consistent with spherical emboli, where
the volume dissolve rate is proportional to the surface area). We take
$\Delta t=1$s. An essential component of the model is an effective
interaction between incoming emboli and pre-existing blockages: at a
bifurcation, emboli are more likely to move in a direction with more
flow.

We use a Monte Carlo simulation to obtain numerical solutions of
the model and determine average behavior. Our algorithm is iterative
with the following steps;
\begin{enumerate}
\item On any time-step, an embolus may be created in the root node of
the tree with probability $\Delta t/\tau$.  In this study, a single
initial embolus size, $d_0$, was assumed.
\item All end nodes are examined to determine if there is a blockage
upstream. This determines the distribution of blockages which are to
be used in the following steps.\label{enum:step2}
\item The emboli move according to the following
rules.

\begin{enumerate} 

\item If the embolus is larger than the current node, it does not move.

\item If all arterioles downstream are blocked, the embolus may not move
since there is no flow.

\item If the embolus is smaller than the current node, and there is
flow downstream,

\begin{enumerate}

\item The proportion of the tree that is blocked downstream from the
embolus to the left, $n_A$, is determined from the distribution of
blockages in step \ref{enum:step2}. The
proportion of unblocked branches in direction $A$ is $X_A = 1- n_A$.

\item Likewise, the unblocked
proportion of branches in direction $B$ is $X_B = 1 - n_B$.

\item The flow proportion in direction $A$ is determined to be $f=X_A/(X_A +X_B)$.

\item The embolus then moves in direction A with probability determined from equation \ref{eqn:leftrightprop}. Otherwise it moves in direction B. 
\end{enumerate}
\end{enumerate}

\item All emboli dissolve, leading to a linear reduction in radius
during each time step. Completely dissolved emboli are removed from
the simulation.

\end{enumerate}

\section{Steady state and limiting behavior}
\label{sec:limits}

We can estimate the total distribution of emboli in the system using a
steady state approximation. The size of an embolus a time $t$ after
introduction to the tree is
%
$d = d_{0} - t\alpha$,
%
where $d_0$ is the initial size of the embolus and $t$ is less than $t_{0} =
d_{0}/\alpha$. Thus, the average number density of emboli with size
$d$ in the vascular system is
$N(d) = \theta(d_0-d)/\alpha\tau$, where $\theta$ denotes a step
function. In clinical situations, there is typically a range of
initial sizes centered about a single size. Simulations indicate that
behavior of the model with a range of initial embolus sizes is
qualitatively similar to that with a single size.

The total number of blockages at a single level is
$N_{m}=\int_{2^{m/\gamma}x_{\rm min}}^{2^{(m+1)/\gamma}x_{\rm min}}N(x)dx$,
%
with the exception of the tree level with similar size to the initial embolus ($2^{m/\gamma}x_{\rm min}<d_0<2^{(m+1)/\gamma}x_{\rm min}$)where
%
$N_{m}=\int_{2^{m/\gamma}x_{\rm min}}^{d_0}N(x)dx$.
%
Choosing a $d_0$ that has the same size as
a node,
\begin{equation}
N_m = x_{\rm min}\left[2^{(m+1)/\gamma} - 2^{m/\gamma}\right] /\alpha\tau ,
\label{eqn:numberblocked}
\end{equation}
assuming the condition $d_{0} > 2^{(m+1)/\gamma} x_{\rm min}$ is
satisfied. If $d_{0} \le 2^{m/\gamma} x_{\rm min}$ there are no emboli
of comparable size to the node and $N_m = 0$. If $d_0$ is not the same
size as at least one of the nodes in the tree, a prefactor
$[d_0 - x_{\rm min}2^{m/\gamma}]/[x_{\rm min}2^{(m+1)/\gamma} - x_{\rm
min}2^{m/\gamma}]$ is associated with $N_{m}$ for the largest set of
nodes with blockages.

Obstructions at different levels of the tree have different consequences for
flow in the end arterioles. For example, an embolus blocking the root
(source) node of the arterial tree prevents blood flow to all end
nodes, whereas a blockage in one of the smallest nodes only affects
flow in that node. In general, the proportion of exit ($m=0$) nodes
not receiving flow because of an obstruction in level $m$ is
$p_m = N_m 2^{m}/N_{\rm last}$. Where $N_{\rm last} = 2^{M_{\ell}-1}$
is the total number of exit nodes in the tree. We also
determine the node level corresponding to blockages by the largest
emboli ($M_{\rm big}$) via $x_{\rm min}2^{M_{\rm big}/\gamma} = d_0$,
so
\begin{equation}
M_{\rm big} = \gamma \log_2(d_0 / x_{\rm min}).
\end{equation}
We have assumed that the initial embolus size is the same as the size of one of the nodes.

In general, the blockage proportion can be determined from the
$n$-point correlation functions, $g_{1\cdots n}$,
\begin{eqnarray}
p &=& \sum_{n=0}^{M_{\rm big}-1}p_n 
- \sum_{n\neq m=0}^{M_{\rm big}-1} g_{nm} p_n p_m + \sum_{l\neq m\neq n=0}^{M_{\rm big}-1} g_{lmn} p_l p_m p_n \nonumber\\
& & + \cdots 
+ (-1)^{M_{\rm big}-1} g_{n_1 ... n_{M_{\rm big}}}\prod_{n=0}^{M_{\rm big}-1} p_n
\end{eqnarray}
To compute all the correlation functions analytically would require
detailed knowledge of the probability for each configuration (an
analogue of the partition function). Since the form for that function
is not known, we examine limiting cases to gain insight. In the low
density limit, there is plenty of space in the tree, and it is
expected that emboli avoid each other (as though there is a strong
interaction). In the very dense limit, space in the tree is limited
and blockages from different emboli have a high probability of
removing flow from the same arterioles, which is indicative of weak
effective interaction.

First, we examine the dilute limit where emboli avoid one another, so
that all emboli block different parts of the tree, i.e. the emboli
cause mutually exclusive halt in the flow as perceived from the exit
nodes. The proportion of the tree that becomes obstructed considering
the steady state configuration is
\begin{equation}
p  = \sum_{m=0}^{M_{\rm big}-1}p_{m} = \sum_{m = 0}^{M_{\rm big}-1} \frac{N_m 2^{m}}{N_{\rm last}},
\label{eqn:exclusive}
\end{equation}
when $\sum_{m=0}^{M_{\rm big}-1} N_{m} 2^{m}/N_{\rm last} < 1$, and
$p=1$ otherwise. The sum ends at $M_{\rm big} -1$ since the node level
corresponding to $M_{\rm big}$ is blocked for an infinitesimally small
length of time before a recently introduced embolus dissolves and moves to the next
layer $M_{\rm big} -1$. Substituting the expressions for $M_{\rm big}$, the sum of the geometric series is,
\begin{equation}
p = \frac{x_{\rm min}[2^{1/\gamma}-1]((d_0 / x_{\rm min})^{(\gamma+1)}-1)}{\alpha\tau 2^{M_{\ell}-1}(2^{(\gamma+1)/\gamma}-1)}.
\end{equation}
This result has potential for clinical significance, as it shows that for
$\gamma>2$ a volume conserving break-up of emboli ($d_0\rightarrow
d_0/2^{1/3}$ and $\tau\rightarrow\tau/2$) leads to a net reduction in
blocked arterioles. As previously discussed, $\gamma\approx 3$ in the vasculature. Our analysis indicates that clot-breaking
therapy could be an important tool for treating acute stroke, provided it
can be delivered quickly. Given that the model is in the early stages of development, we add the caveat that this result is merely indicative and should not be used directly for clinical applications.

For a state where the positioning of emboli in all layers is
independent (non-interacting problem) the total probability of end
arterioles being blocked is reduced by the overlap between co-existing
blockages in different layers,
\begin{equation}
p = p_0 + (1 - p_0)p_1 + (1 - [p_0 + (1 - p_0)p_1])p_2 + \cdots.
\label{eqn:blockexplain}
\end{equation}
The first term in Eq. \ref{eqn:blockexplain} is the probability of an embolus in level 0 stopping
flow to end arterioles. Emboli blocking the next layer have a
probability $p_1$ of blocking the remaining $(1-p_0)$ arterioles still receiving flow after the level 0 blockages are taken into account  and so
on. Eq. \ref{eqn:blockexplain} may be rewritten
\begin{eqnarray}
p &=& \sum_{n=0}^{M_{\rm big}-1}p_n 
- \sum_{n\neq m=0}^{M_{\rm big}-1} p_n p_m\label{eqn:independent}\\
& &+ \sum_{l\neq m\neq n=0}^{M_{\rm big}-1} p_l p_m p_n + \cdots 
+ (-1)^{M_{\rm big}-1} \prod_{n=0}^{M_{\rm big}-1} p_n.\nonumber
\end{eqnarray}

There is a special value for the parameters in
Eq. \ref{eqn:independent} where no flow reaches end arterioles
($p=1$). This occurs when the level with the largest emboli (of size
$d_0$) becomes fully saturated. This is because more emboli block the
largest nodes as shown by Eq. \ref{eqn:numberblocked} and the largest
nodes supply the most arterioles. Examining the $p=1$ case leads to an
equation that relates the parameters of the system when it is just
fully blocked,
%
$N_{M_{\rm big}-1} 2^{M_{\rm big}-1}/N_{\rm last} = 1$,
%
which leads to a prediction of the embolus size corresponding to all
end nodes blocked,
\begin{equation}
d_c =  \left[ 2^{(\gamma + 1) / \gamma} \alpha\tau 2^{M_{\ell}-1} x_{\rm min}^{\gamma} / (2^{1/\gamma}-1) \right]^{1/(\gamma + 1)}.
\end{equation}
It is interesting to note that since $d_c^{\gamma+1}\propto \tau$ this
result leads to the same conclusion about clot busting as in the
dilute (strong-interacting) limit, i.e. that breaking up emboli
reduces blockage. We note that time dependent fluctuations about the
average density have not been considered in our analysis. Such
fluctuations slightly increase the embolus size required to completely
block the tree, but become irrelevant if the correct scaling to an
infinite lattice is made.  The point at which the tree becomes fully
blocked is similar to a percolation threshold, since there are no
direct paths from the root node to the exit nodes (arterioles).

We note that our findings for the interacting and non-interacting
limits indicate quite different behavior. For the low density (strong
interaction) limit, emboli avoid each other. For the very dense (weak
interaction) limit, space in the tree is limited and blockages from
different emboli have a high probability of removing flow from the
same arterioles. As shown by our analytic results, the two limits
exhibit quite different dependence on embolus prevalence (which depends
on embolization rate, dissolve rate and size). Eq. \ref{eqn:exclusive}
for the dilute limit corresponds to all
$g=0$. Eq. \ref{eqn:independent} corresponds to all $g = 1$. Thus, we
expect either a sharp transition or crossover behavior between the
dilute and dense limits.

\section{Time correlator}
\label{sec:timecorrelator}

\begin{figure}
\includegraphics[height=75mm,angle=270]{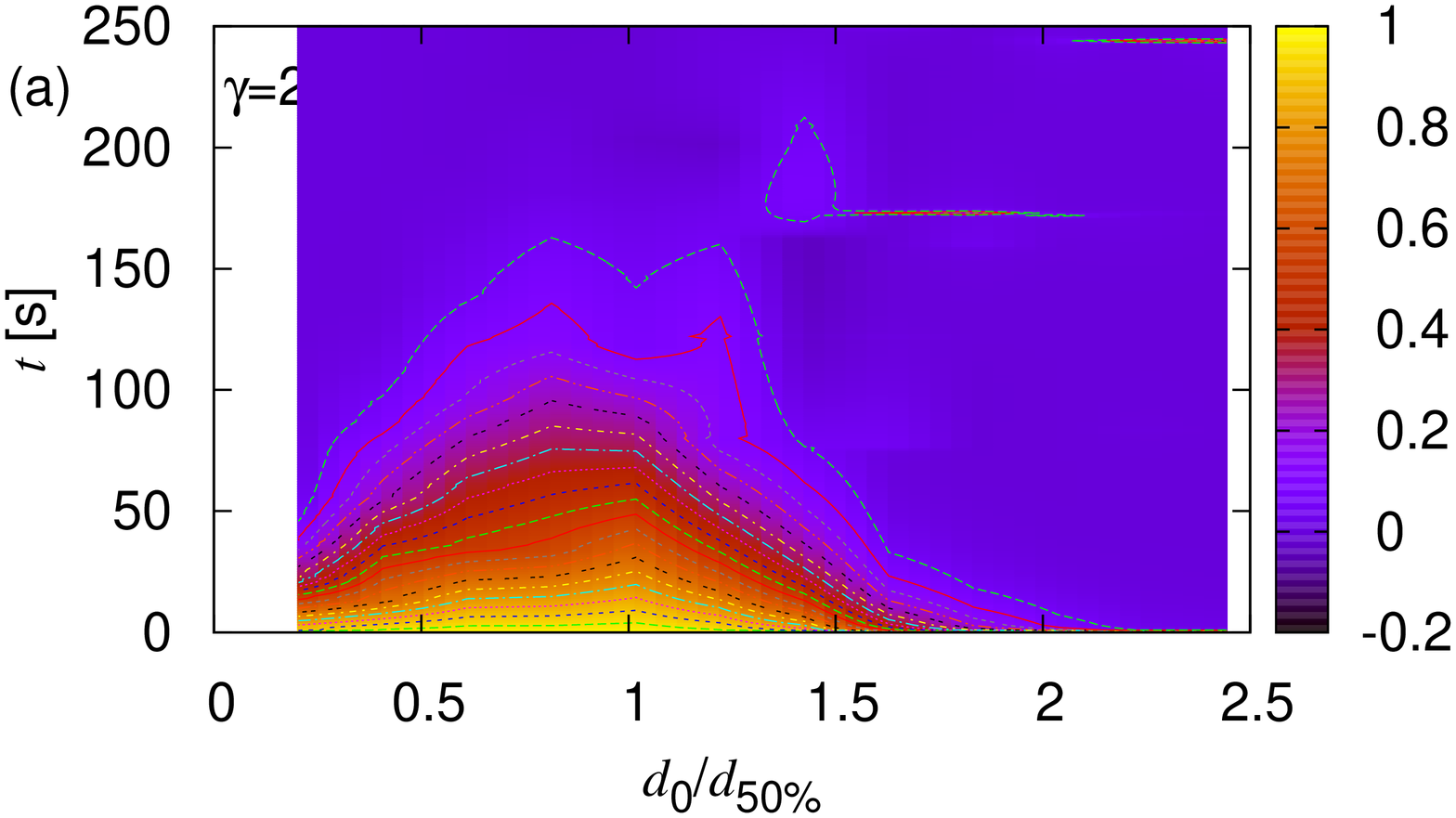}
\includegraphics[height=75mm,angle=270]{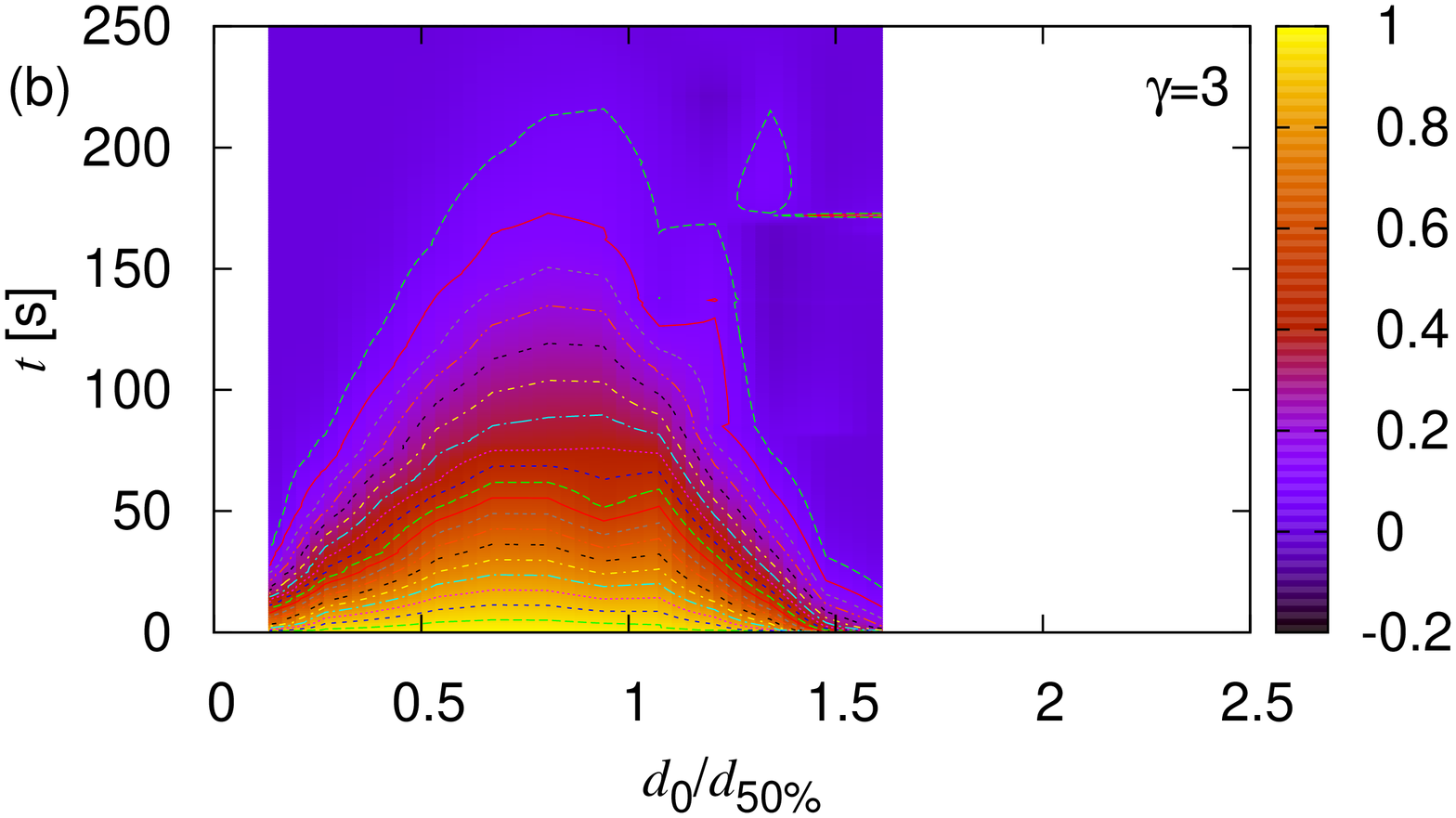}
\includegraphics[height=75mm,angle=270]{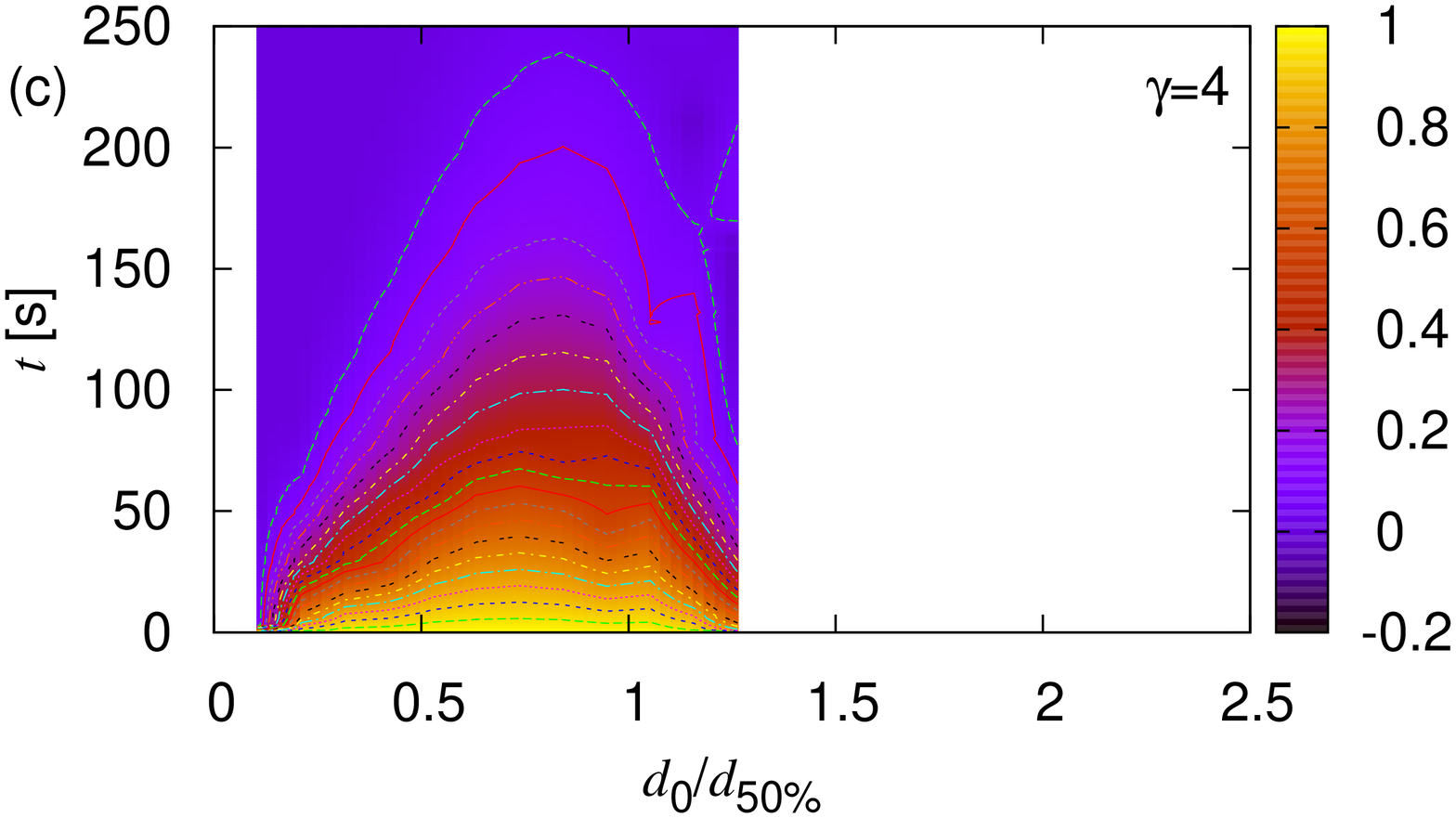}
\caption{(Color online) Finite-size scaling showing the change in the form of the time autocorrelation function $G(t)$ with varying gamma: (a) $\gamma=2$, (b) $\gamma=3$, (c) $\gamma=4$. Simulations are carried out using the linear trajectory weighting. The tree had 20 levels, and $\alpha=3\times 10^{-4}$ and $\tau=0.1$. The lines are contours on $G(t)$. On increasing $\gamma$, the timescale can be seen to increase, with the shape of the contours becoming less rounded.}
\label{fig:tacgammalinear}
\end{figure}

\begin{figure}
\includegraphics[height=75mm,angle=270]{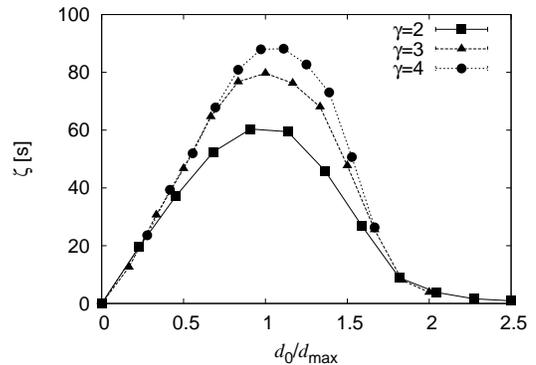}
\caption{The curves of integrated correlation time become more sharply peaked on increasing $\gamma$. $\tau=0.1$, $\alpha=3\times 10^{-4}$. Curves are shown for various $\gamma$ and linear trajectory weighting. The tree had 20 levels, and $\alpha=3\times 10^{-4}$ and $\tau=0.1$. $d_{\max}$ is the maximum in the integrated correlation time curve.}
\label{fig:correlationtime}
\end{figure}

\begin{figure}
\includegraphics[height=75mm,angle=270]{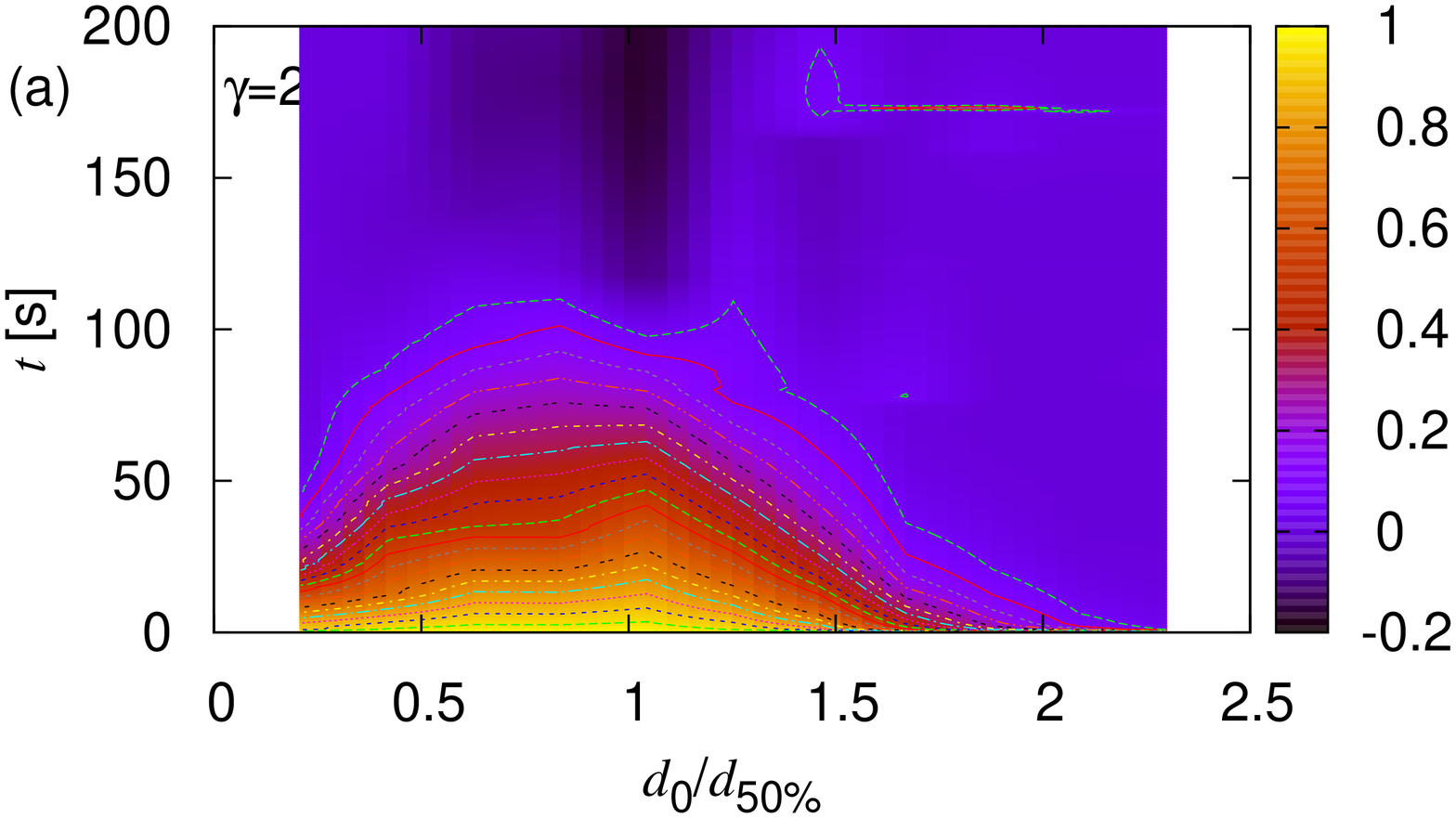}
\includegraphics[height=75mm,angle=270]{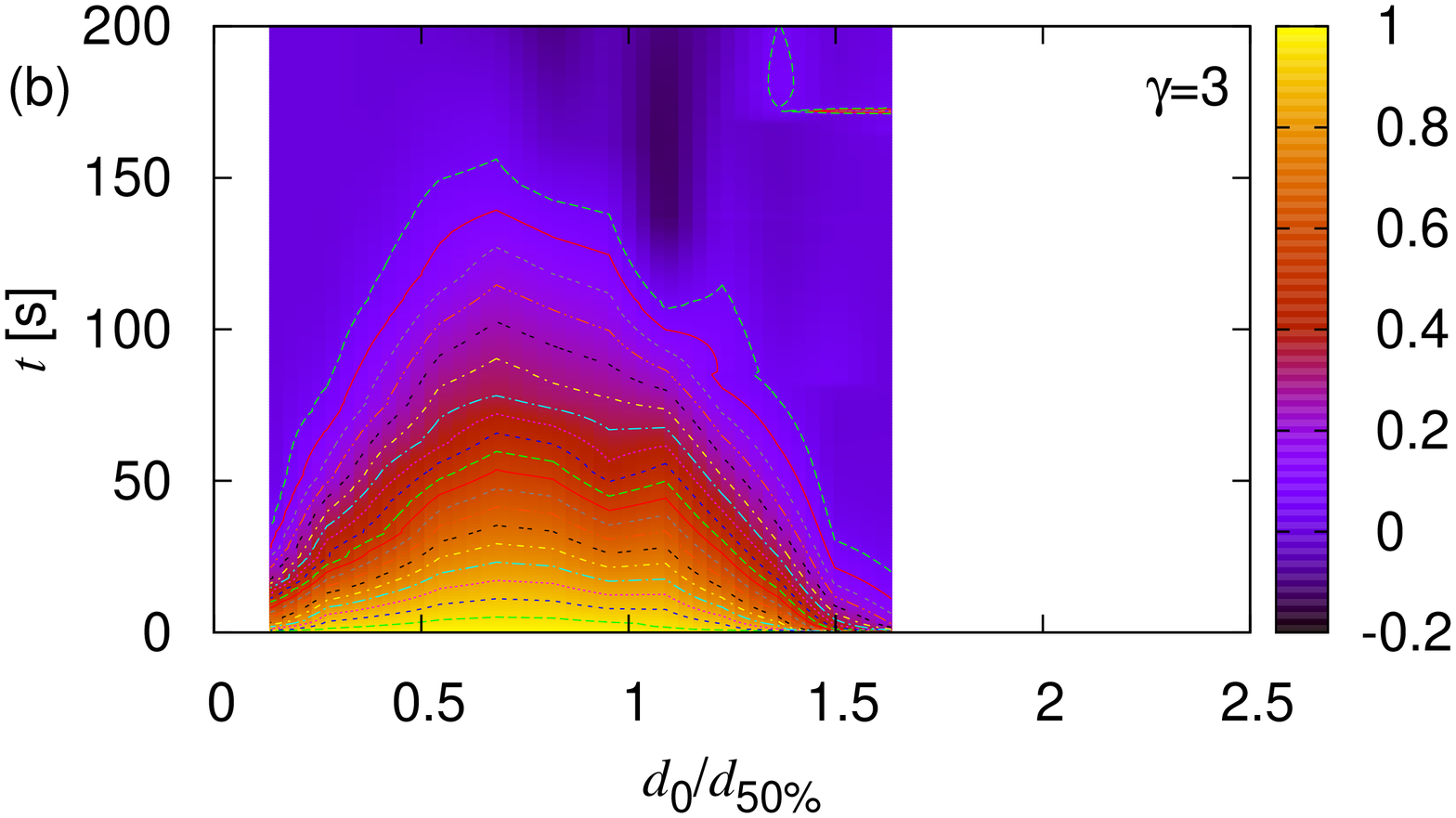}
\includegraphics[height=75mm,angle=270]{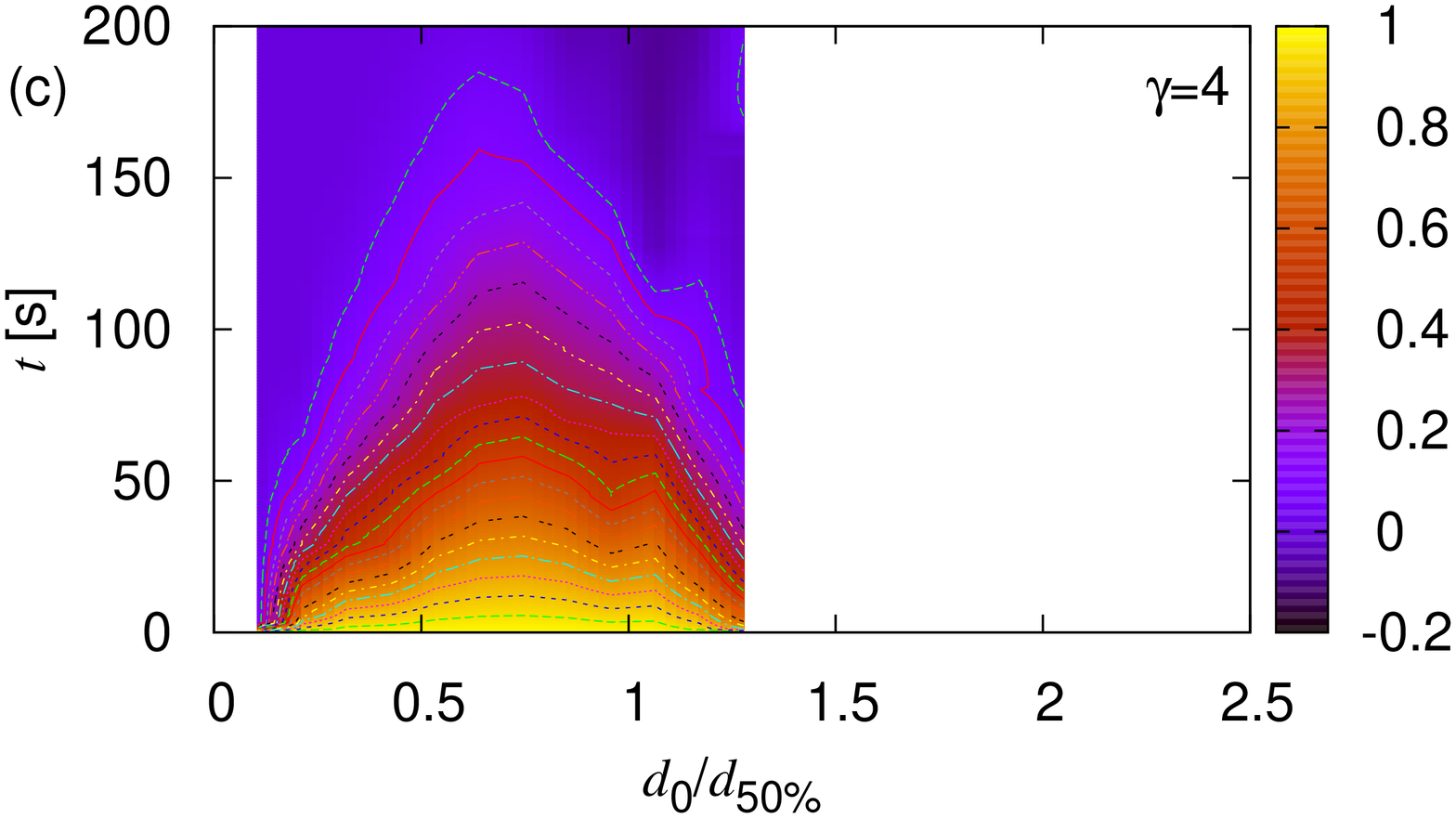}
\caption{(Color online) Change in time autocorrelation function $G(t)$ with varying gamma: (a) $\gamma=2$, (b) $\gamma=3$, (c) $\gamma=4$. Simulations are carried out using the step function trajectory weighting. The tree had 20 levels, and $\alpha=3\times 10^{-4}$ and $\tau=0.1$. The time correlations become significantly longer and the peak in the contours becomes rapidly sharper as $\gamma$ increases.}
\label{fig:tacgammastep}
\end{figure}

A key question in this article is whether there is crossover or
transition between the dilute and dense limits. In a non-equilibrium
system at criticality, the timescale associated with fluctuations
would diverge (in this case, the length of time that an additional
embolus added to the system contributes to an overall increase in the
level of obstruction). However, if there is a crossover, then while
there could be increase in the timescale of correlations, no
divergence would be seen.
The appropriate measure of such fluctuations is the time autocorrelation function, which is computed from the
expression,
\begin{equation}
G(t) = \frac{\langle(p(t)-\bar{p})(p(0)-\bar{p})\rangle}{\langle(p(0)-\bar{p})^2\rangle}
\end{equation}
and the time correlation function according to,
\begin{equation}
\Gamma(t) = \frac{1}{N_0}\sum_{i=1}^{N_0}\langle(p_i(t)-\bar{p})(p_i(0)-\bar{p})\rangle,
\end{equation}
where $p_{i}(t)=1$ if end node $i$ is blocked at time $t$ and $p_{i}(t)=0$ otherwise.

In a critical system, $\Gamma(t)$ should have the form,
\begin{equation}
\Gamma(t) \propto \frac{1}{t^{y}}e^{-t/\zeta}
\end{equation}
where $y$ is a constant, and $\zeta$ is the correlation time. If
$G(t)$ is positive at all times (i.e. there is no anticorrelation) it
is possible to estimate the correlation time by summing (integrating)
over the autocorrelation function, $\zeta\approx\Delta
t\sum_{n=0}^{\infty}G(n\Delta t)$ (this measure is commonly known as
the integrated correlation time).

To examine the existence of a phase transition, it is necessary of investigate the behavior of the system as the size is increased (note that the finite size scaling of the order parameter is discussed in the next section). The correct way of increasing the effective size of the system is unconventional. It is necessary to
make the size difference between vessels smaller, which means that the
emboli interact with more levels as they dissolve \footnote{Note that simply decreasing the rate at which the emboli dissolve, while keeping the
number of emboli in the system constant (by decreasing the
introduction rate of emboli) is not sufficient.}. This can be done by
increasing $\gamma$, since the difference in vessel sizes between
levels is $2^{1/\gamma}-1$. Fig.  \ref{fig:tacgammalinear} shows the
change in time autocorrelation function $G(t)$ with varying $\gamma$,
when the linear trajectory weighting is used. So that the simulations
with different $\gamma$ can be compared, we use the dimensionless
parameter $d_0/d_{50\%}$. The size $d_{50\%}$ corresponds to half of
end nodes receiving no flow. We note that $d_{50\%}$ is approximately
proportional to $\gamma$. A distinct change in the shape of the
contours can be seen, with the time scale increasing most rapidly
towards the center of the graph as $\gamma$ is increased. This can be
examined in the context of the integrated correlation time.

Figure \ref{fig:correlationtime} shows Monte Carlo results for the
integrated correlation time. 
Increase in $\gamma$ has an effect on the correlation time consistent
with a phase transition, with the hump associated with the maximum of
the curve becoming more defined, and moving closer to the form
consistent with a divergence, and thus a phase transition, for large $\gamma$. Time and
memory constraints limit the maximum size of the tree, so we have been
unable to examine systems with $\gamma>4$.

Finally, the time auto-correlators are examined in the context of the
step function weighting of embolus trajectory, with the results shown
in Fig. \ref{fig:tacgammastep}. Again an increase in the timescale is
seen on increasing $\gamma$, with qualitatively similar changes in the
form of the graphs to those seen for the linear weighting scheme. In
the case of the step function weighting, there are regions of
anti-correlation, so the integrated correlation time can not be used
to examine the timescale. Since there is evidence of a phase
transition at large $\gamma$, it remains to find a plausible candidate for the order
parameter.

\section{Space-time plots and order parameter}
\label{sec:spacetime}
\label{sec:orderparameter}

\begin{figure}
\includegraphics[width=60mm]{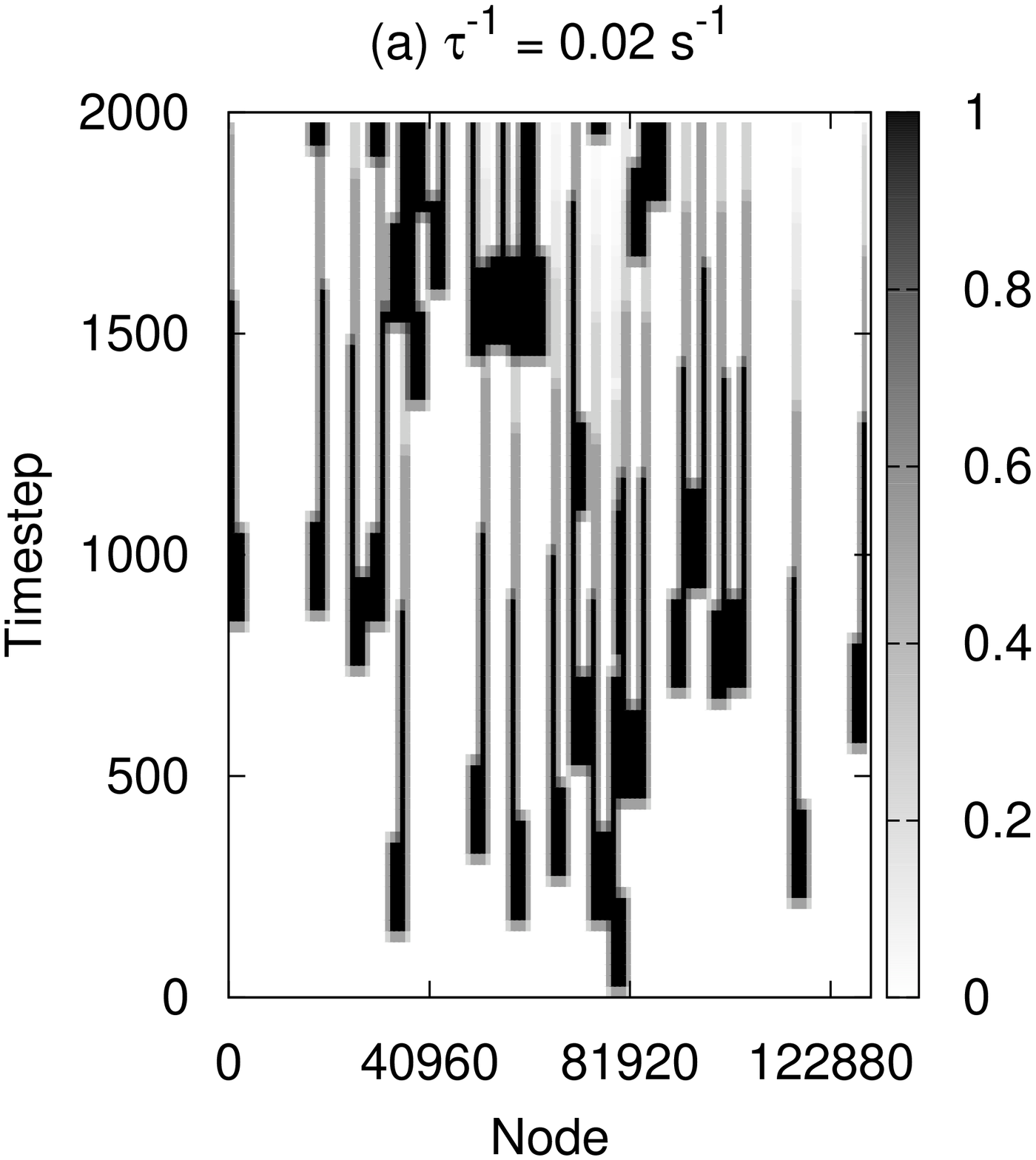}
\vskip -15mm
\includegraphics[width=60mm]{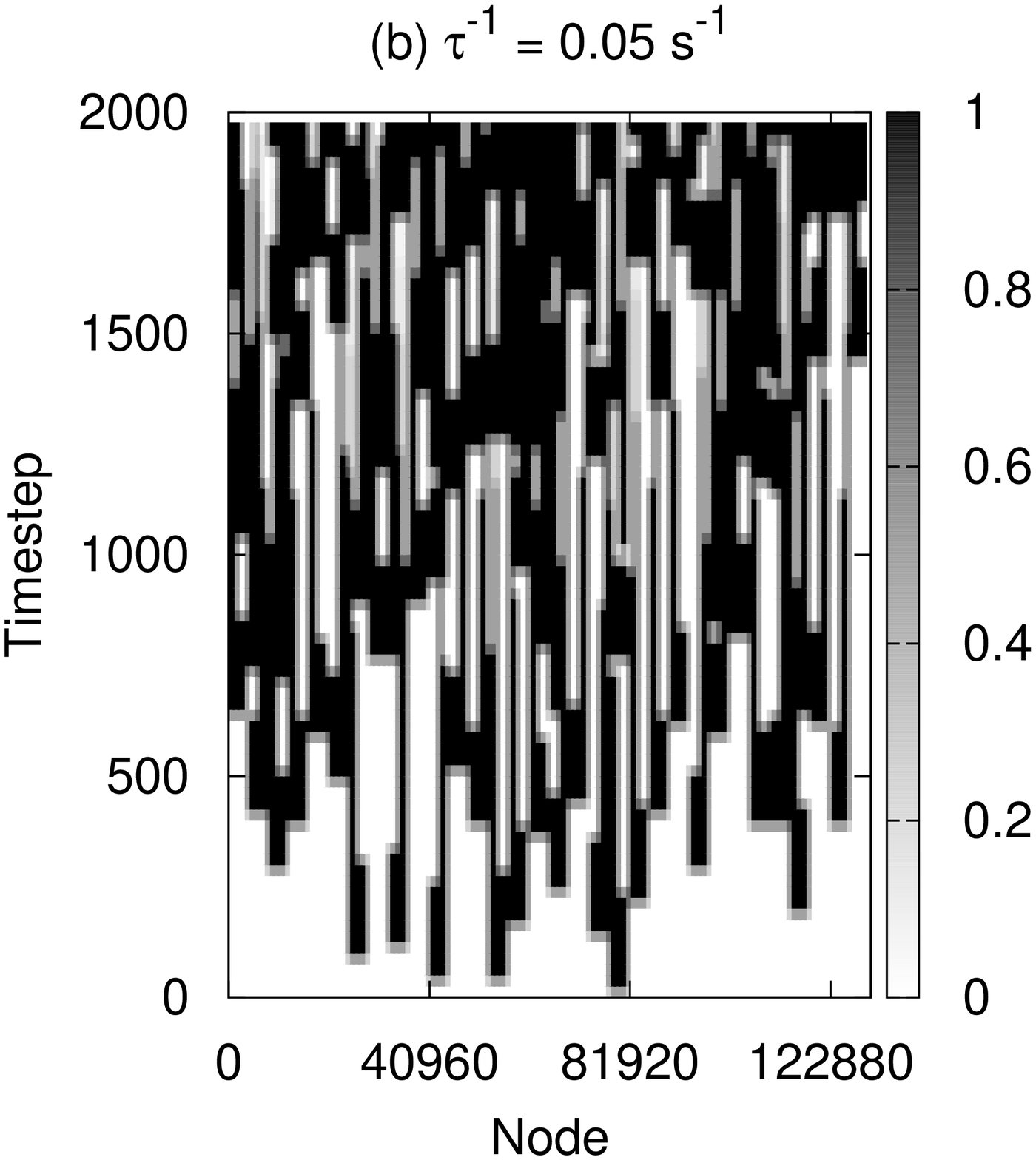}
\caption{Space-time plot showing flow in end arterioles in the model with $b=1$ (linear weighting model). The tree
consists of $M_{\ell} = 18$ levels with $\gamma = 2$, $d_0=0.2$mm and
$\alpha = 10^{-4}$mms$^{-1}$. Fully blocked regions are shown in
black, and free flow in white. Panel (a) shows the dilute limit where
emboli have available space to avoid each other. In contrast (b) shows
the dense limit, where emboli are forced to occupy the same trees
since there is no remaining space. This indicates that overlapped blockages are relevant to the phase transition.}
\label{fig:history}
\end{figure}

\begin{figure}
\includegraphics[width=60mm]{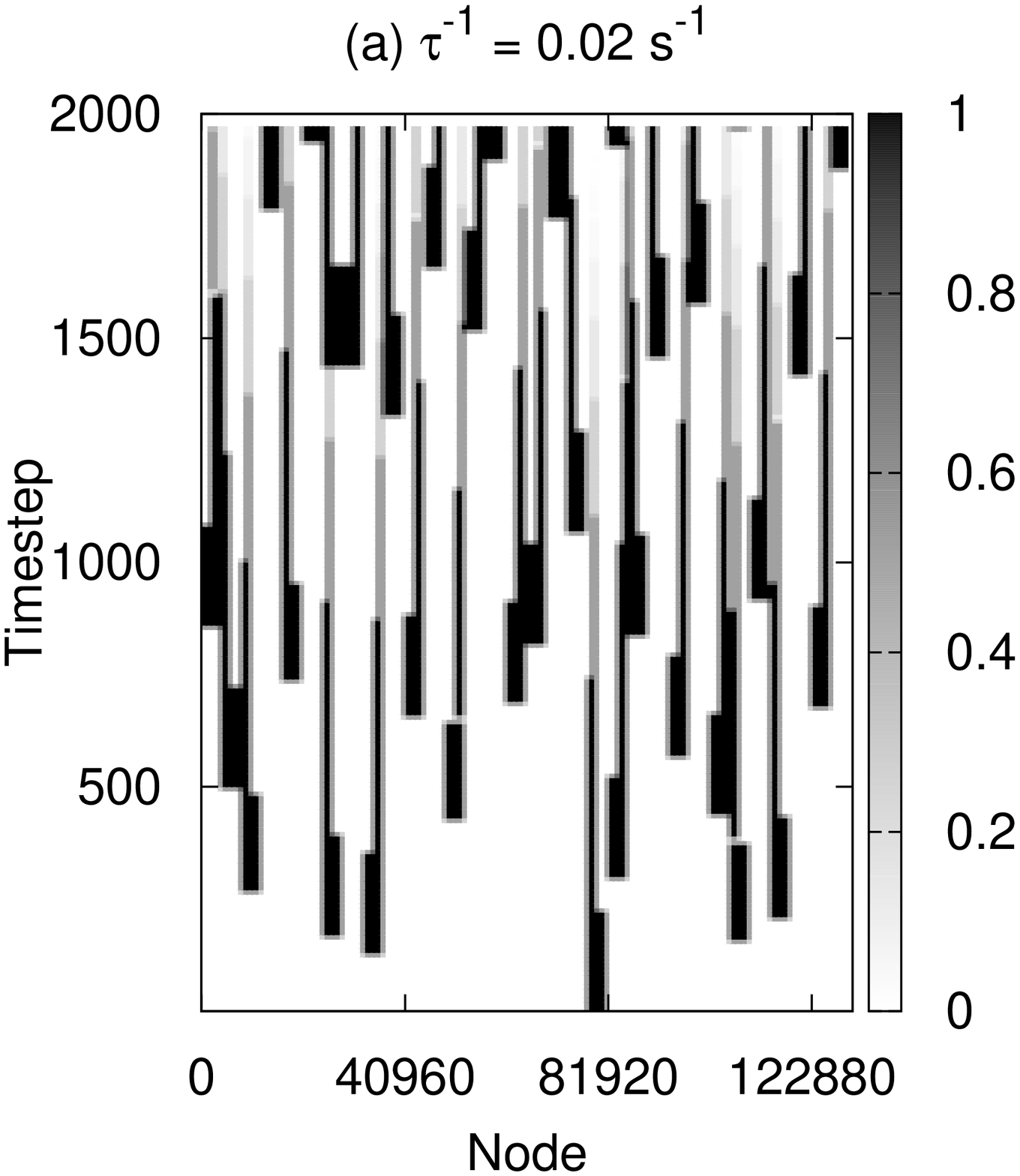}
\vskip -15mm
\includegraphics[width=60mm]{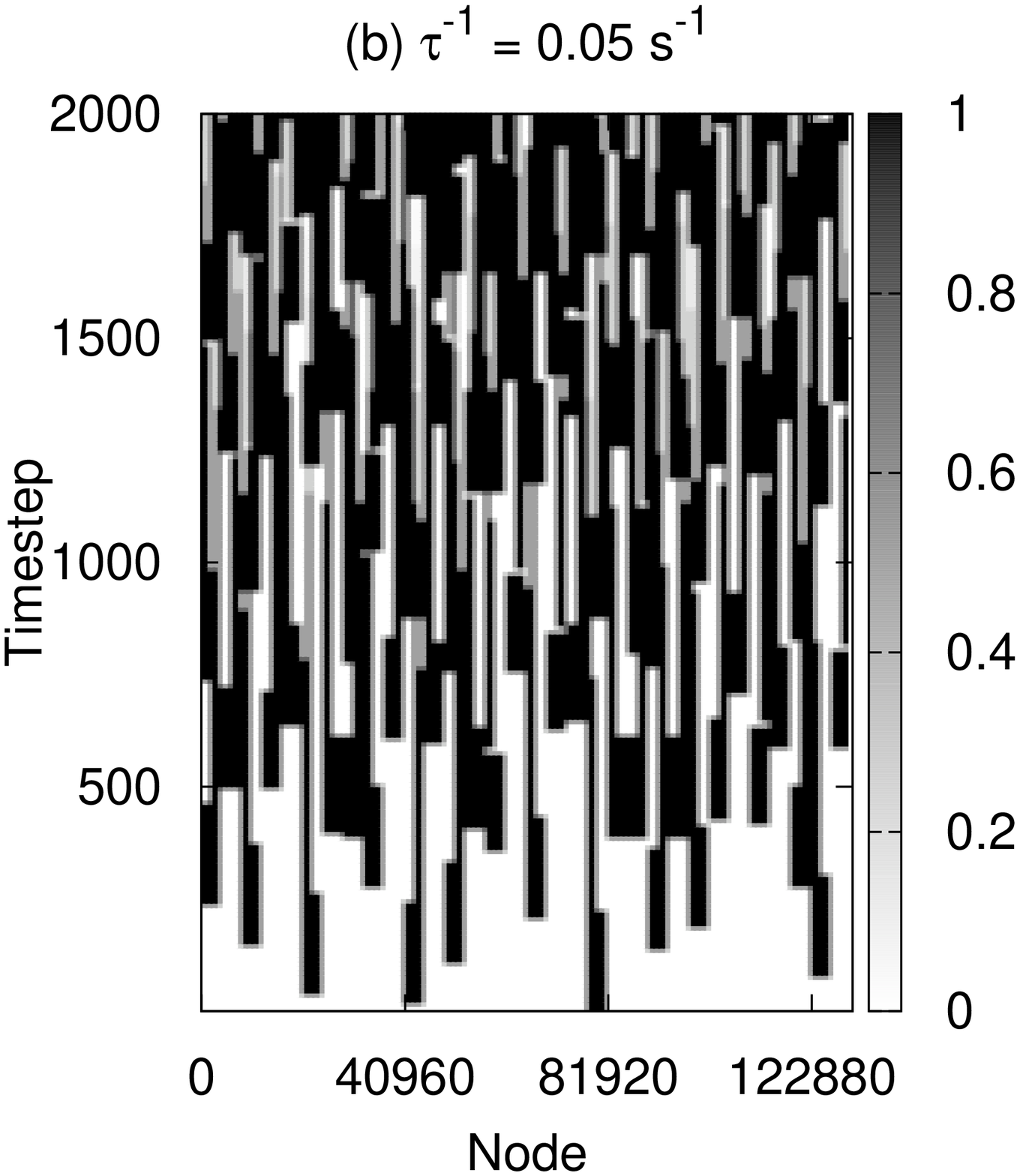}
\caption{As Fig. \ref{fig:history} but with
$b\rightarrow\infty$ (step function trajectory weighting). The emboli become more evenly spaced in the tree, leading to a state that appears to be more ordered in space.}
\label{fig:historymod}
\end{figure}

\begin{figure}
\includegraphics[height=78mm,angle=270]{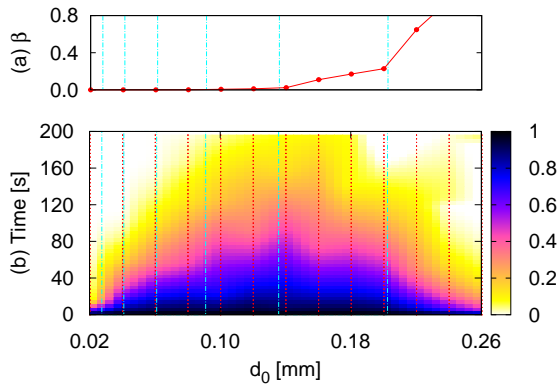}
\caption{(color online) Main panel: Time correlator as $d_0$ is
changed. $M_{\ell}=20$, $\gamma = 1.736$, $\tau^{-1} = 0.1$s$^{-1}$
and $\alpha = 3\times 10^{-4}$mm/s. Errors on individual points are
approximately 3\% of the $t=0$ value (3 standard deviations). The
timescale is significantly increased close to the transition. Top panel: The overlap between flow
shadows shows how the significant increase in timescale is associated
with the onset of order.}
\label{fig:timecorrelator}
\end{figure}

\begin{figure}
\includegraphics[width=78mm]{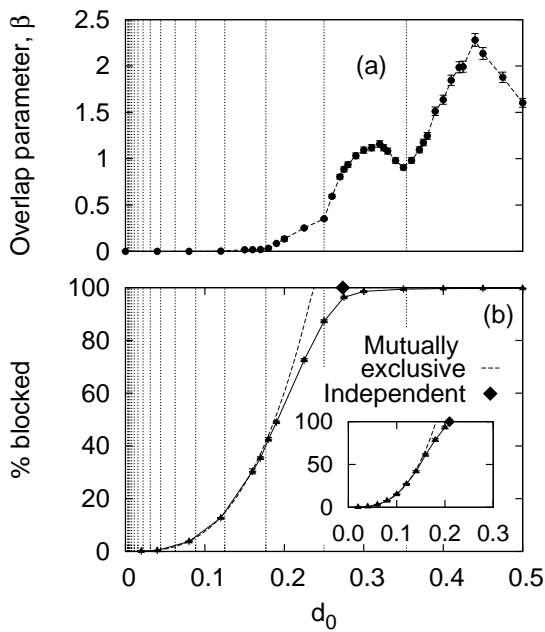}
\caption{(a) Overlap parameter, $\beta$ showing a second-order
transition. The hump features are related to partially occupied levels. (b) Percentage of blocked states and limiting behavior. The vertical lines
indicate the sizes of nodes at each level of the tree. $M_{\ell}=18$ with
$\gamma = 2$, $\tau^{-1} = 0.1$s$^{-1}$ and $\alpha = 3\times
10^{-4}$mm/s. (Inset: $M_{\ell} = 20$, $\gamma=2.151$, $\tau^{-1}=0.1$s$^{-1}$ and $\alpha = 10^{-4}$mm/s.) Where no error bars are visible, they are smaller than
the markers. For
these parameters, overlap between blockages suddenly increases at
$d_0=0.18$mm in a second order transition. The percentage
of blocked end nodes is always non-zero. }
\label{fig:orderparam}
\end{figure}

\begin{figure*}
\includegraphics[height=75mm,angle=270]{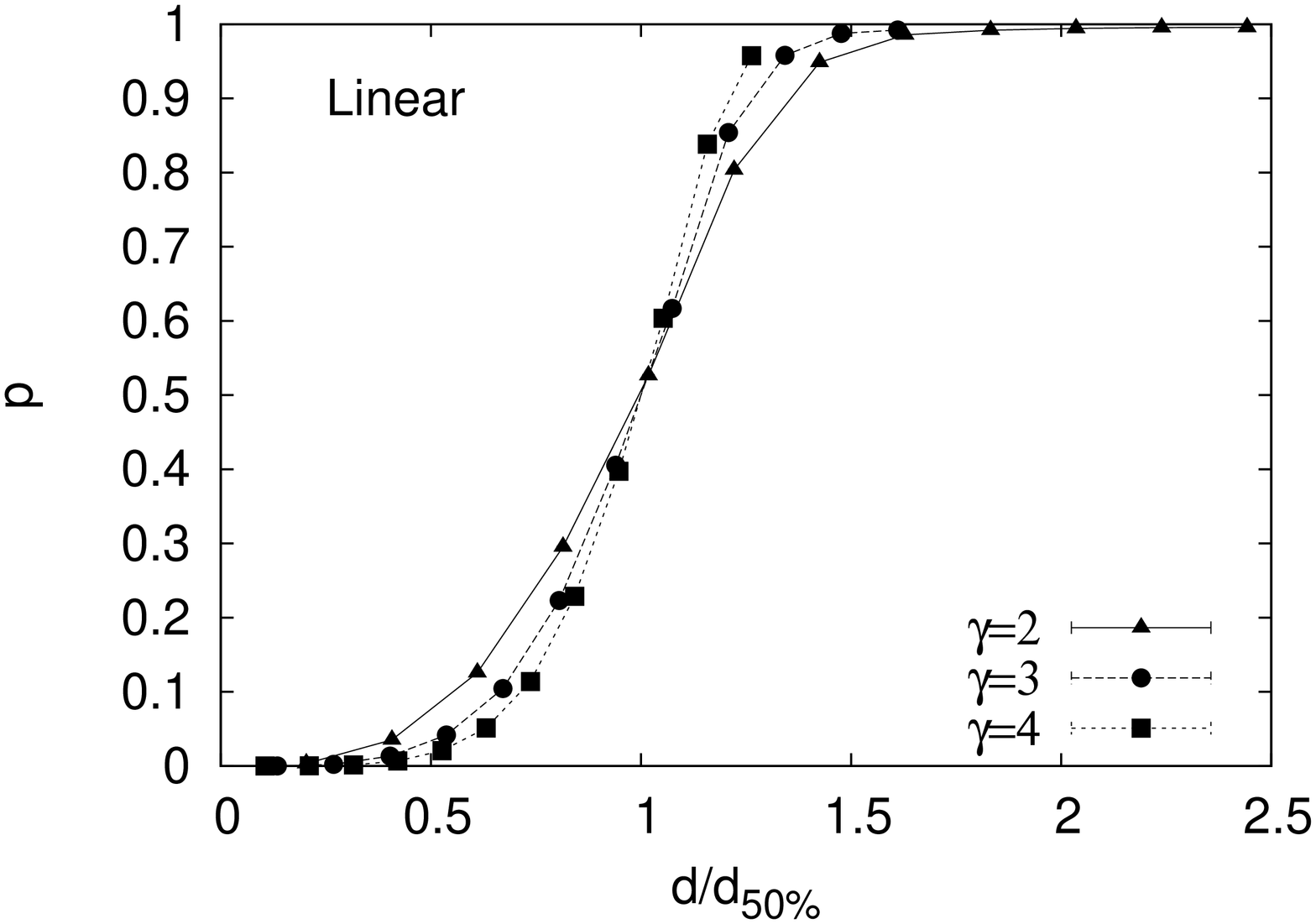}
\includegraphics[height=75mm,angle=270]{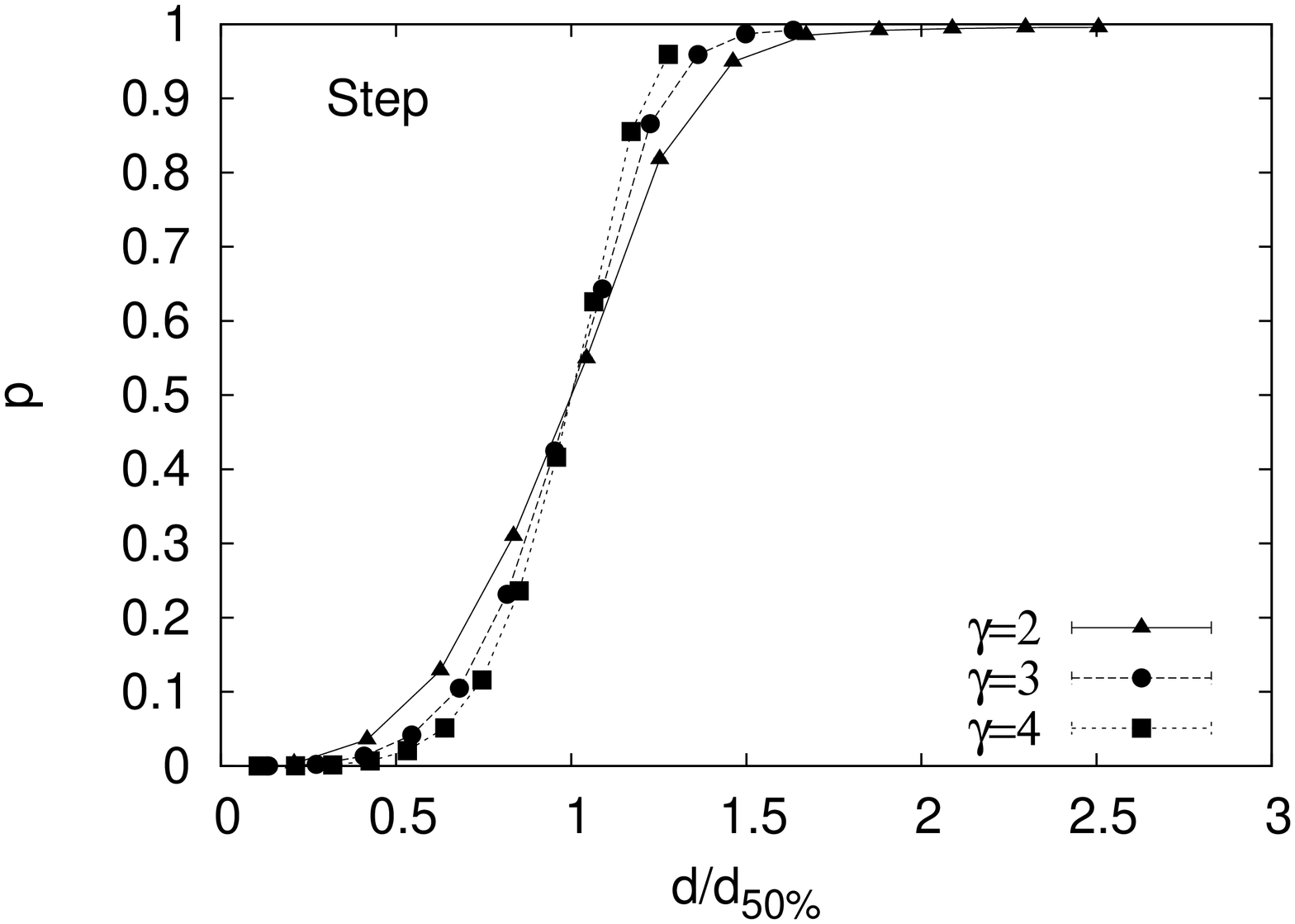}
\includegraphics[height=75mm,angle=270]{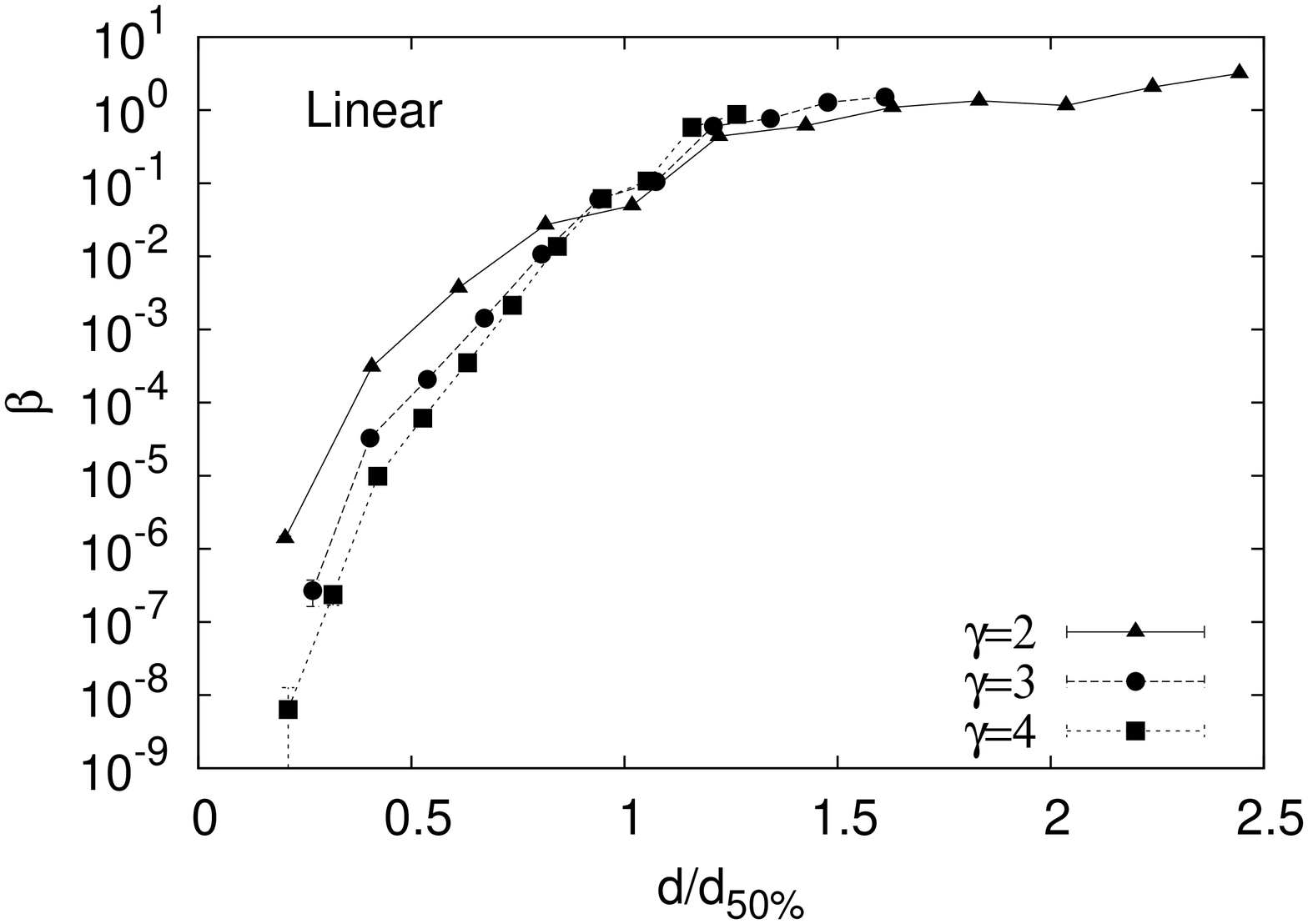}
\includegraphics[height=75mm,angle=270]{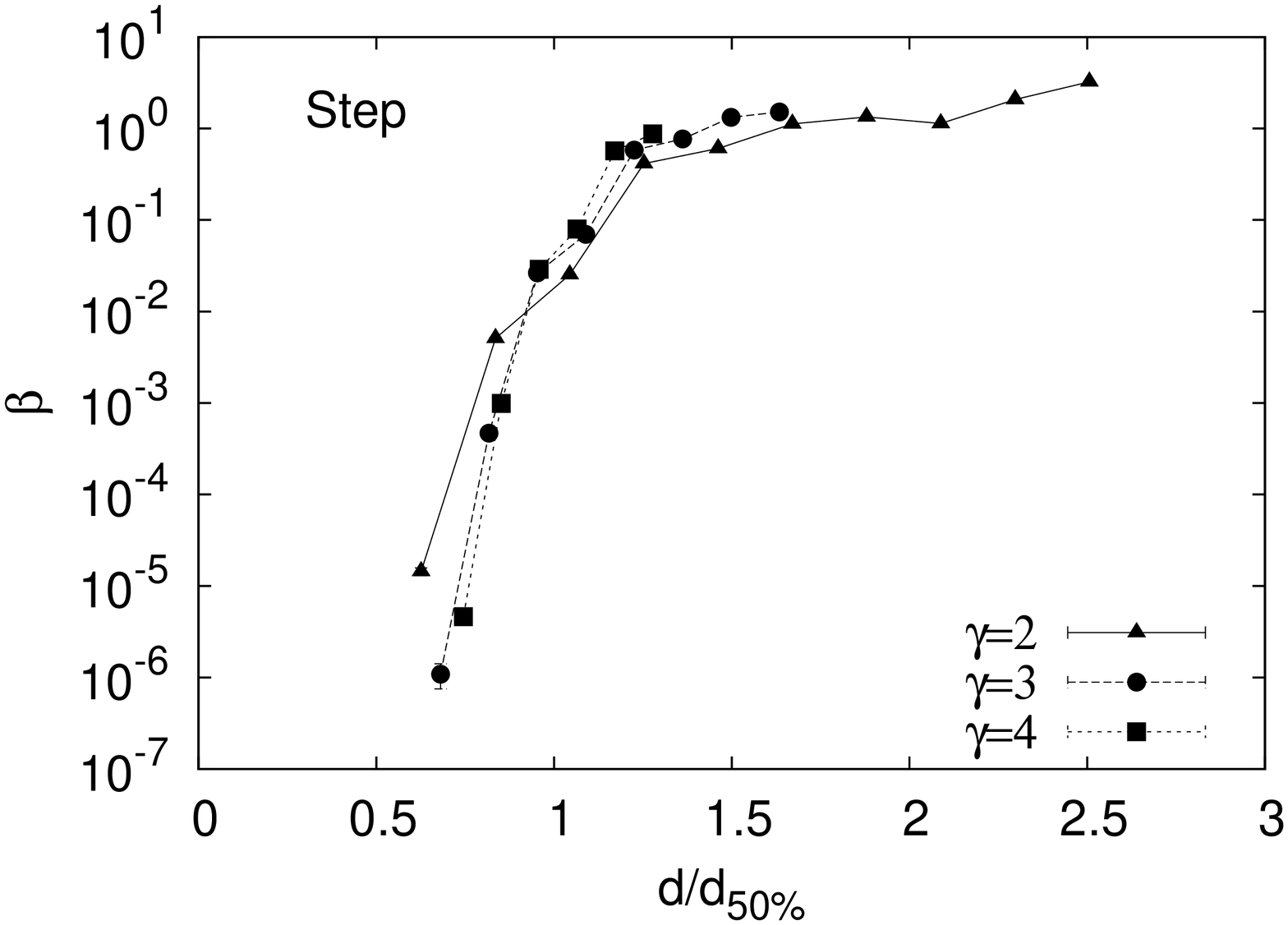}
\caption{Blockage and overlap parameter as $\gamma$ is varied to carry out the finite size scaling. Simulations are carried out using both step function and linear trajectory weighting. The tree had 20 levels, $\alpha=3\times 10^{-4}$ and $\tau=0.1$. Some of the small $d_0$ points are missing on the plots relating to the step function trajectory weighting, since $\beta=0$. As $\gamma$ is increased, the values of $\beta$ decrease for $d_0<d_{50\%}$ and $\beta$ increases for initial embolus sizes $d_0>d_{50\%}$. Scaling of this type is consistent with a phase transition.}
\label{fig:finitesize}
\end{figure*}

In order to examine how the model changes on going through the
transition, we followed the motion of emboli through the model
arterial tree for specific realizations of the
embolization. Fig. \ref{fig:history} displays spacetime plots showing
which capillaries (end nodes) do not receive flow during a single
run when the linear flow
weighting scheme was used. Panel (a) shows the dilute limit where
emboli are able to avoid each other. Each `wedge-like' pattern represents the lifetime of an embolus, which initially blocks a large number of nodes, then dissolves and moves to block a smaller node. There is no clear spatial ordering between the emboli. In contrast (b) shows the dense
limit, where space limitations force emboli to cast `flow shadows' on
the same arterioles: emboli in different flow levels cause the same
end nodes to lose flow. This indicates that overlapping blockages
could be important in the transition.

It is also of interest to see what differences the step function weighting scheme makes to the distribution of emboli in the vasculature. Fig. \ref{fig:historymod} shows spacetime plots for
$b\rightarrow\infty$ (step function trajectory weighting) with model parameters that are otherwise the same as for Fig. \ref{fig:history}. The effect of the step function weighting is to strongly redirect the emboli away from blocked vessels. Thus the emboli
become more evenly spaced in the tree, leading to a state that appears
to be more ordered in space. Again, when large numbers of emboli are present in the network, the tree becomes sufficiently populated that emboli can no longer avoid each other, so the flow shadows from blockages overlap.

The spacetime plots in Figs. \ref{fig:history} and \ref{fig:historymod} indicate that the main change in the state of the model on increasing embolus size is the overlap between blockages. Therefore, we suggest that the following measure of the overlap between flow shadows, $\beta$, may act as an order parameter,
\begin{equation}
\beta = \langle \sum_{i\ne j}n_{i}n_{j} \rangle,
\end{equation}
where
$n_i = 1$ if an embolus in level $i$ stops flow in the end arterioles
at level $0$.

If $\beta$ were the order parameter, then the timescale associated with correlations would peak at the embolus size where the order parameter becomes finite. To investigate this, Fig. \ref{fig:timecorrelator} shows Monte Carlo results revealing how the
time autocorrelation function changes with initial embolus size (bottom panel) alongside the overlap parameter (top panel). The hump features in both $\beta$ and the time autocorrelation function are
related to prefactors introduced when the largest emboli do not have
the same size as a node. The timescale
associated with correlations is largest when $d_0=0.14$mm, corresponding to the embolus size where overlap of flow shadows becomes significant. This is a strong indicator that $\beta$ is the order parameter.

We also address whether the proportion of blockages could be the order parameter, rather than the overlap of blockages. Fig. \ref{fig:orderparam}(a) shows numerical results for the overlap
of blockages, $\beta$. For these parameters, overlap between
co-existing blockages suddenly increases at $d_0=0.18$mm. In contrast,
panel (b) shows that the percentage of blocked end nodes is always non-zero and is unlikely to be the
order parameter of the non-equilibrium transition. The limiting behaviors are also displayed on panel (b) for completeness and agree with the Monte Carlo results. Generally, the steady state approximation becomes more
appropriate as system size increases.

We complete this section by examining the blockage $p$ and overlap parameter $\beta$ as $\gamma$ is varied to carry out the finite size scaling, as shown in Fig. \ref{fig:finitesize}. We reiterate that the finite size scaling associated with this problem is unusual. Increasing $\gamma$ means that the emboli (which dissolve linearly with time) interact with increasing numbers of nodes. As $\gamma$ is increased, the change in the proportion of blocked end nodes on increasing $d_0/d_{50\%}$ is more abrupt. As we showed in the last section, the timescale associated with fluctuations becomes more sharply peaked when $\gamma$ increases. We also show the overlap parameter $\beta$ on a logarithmic scale. $\beta$ changes over several orders of magnitude as $d_0$ is changed. As $\gamma$ is increased, the values of $\beta$ decrease for $d_0<d_{50\%}$ and $\beta$ increases for initial embolus sizes $d_0>d_{50\%}$. Scaling of this type is consistent with a phase transition, and can be seen for both the step-function and linear trajectory weighting. 

\section{Summary and conclusions}
\label{sec:summary}

We have discussed the statistical physics of cerebral embolization
leading to stroke based on analysis of a minimal model of particles
moving through the cerebral arteries. We investigated whether the
onset of stroke is associated with a gradual crossover or a phase
transition. Our model is non-trivial since the trajectories of emboli
are weighted away from blocked arterioles, generating an effective
interaction. Examination of the time correlator showed that the
timescale of fluctuations increases significantly at a critical value,
consistent with a phase transition. Finite size scaling adds further
weight to this view. We note that finite size scaling shows that the
phase transition strictly occurs when the bifurcation exponent
$\gamma=\infty$, since for reasons discussed earlier in this article
the system has a finite size when $\gamma$ is finite. Thus crossover
behavior is found when $\gamma$ is finite, but is controlled by the
large $\gamma$ phase transition. The order parameter was identified as
the overlap of blockage flow shadows defined by $\beta = \langle
\sum_{i\ne j}n_{i}n_{j} \rangle$. Our model shows that the onset of
stroke can be thought of as driven by non-equilibrium critical
behavior.  Further modeling incorporating increasingly realistic
embolus parameters and anatomy promises to aid our understanding of
embolic stroke, cerebrovascular disease and perfusion injury.

We briefly summarize the differences and similarities between our work
and the directed percolation problem \cite{hinrichsen2000a}. In the
directed percolation problem, flow stops when the number of broken
bonds reaches a critical value corresponding to an absence of routes
from source to exit. There are some crucial differences between the
standard directed percolation problem, and the one described here. The
first major difference is that the liquid is carrying particles which
can break bonds and move the system closer to the percolation
threshold. The second difference is that flow routes re-form after a
characteristic time as emboli dissolve. The third crucial difference
is that the properties of the tree downstream from the input influence
the direction in which bond-breaking particles are carried, so
blockages are not randomly distributed. In this respect, our model
represents an entirely new type of percolation problem.

The existence of a phase transition in the flow could be of clinical
relevance (although we add the caveat that our results should not be
directly applied to medical treatment at the current time). Not only
would an organized state have more blocked nodes than a randomly
ordered positioning of emboli in the cerebral arteries, but also the
increased time scales associated with fluctuations close to the
transition would tend to increase the risk of brain damage during
embolization.

Further work will be carried out to improve the model. For example,
all branchings are currently symmetric, whereas asymmetric branchings
are more common in the vasculature, so an algorithm will be developed
to construct more realistic trees. Also, we are currently carrying out
detailed laboratory tests of the paths of emboli at single asymmetric
bifurcations, and in a silica replica of the major cerebral arteries,
and these results will guide realistic development of the
computational model.

\section{Acknowledgments}

JPH acknowledges useful discussions with Richard Blythe, Michael
Wilkinson and Andrey Umerski. EMLC acknowledges the support of the
British Heart Foundation.

\bibliography{strokecritical_clean_PRE}

\end{document}